\documentclass[letterpaper,twocolumn,10pt]{article}
\usepackage{usenix2019_v3}

\usepackage[utf8]{inputenc}
\usepackage{times}
\usepackage{hyperref}
\hypersetup{pdfstartview=FitH,pdfpagelayout=SinglePage}

\usepackage[printwatermark]{xwatermark}
\usepackage{xcolor}
\usepackage{graphicx}
\usepackage{lipsum}

\usepackage{graphicx, xspace, outlines,url,grffile,enumerate,listings,appendix,algorithm}
\usepackage[normalem]{ulem}
\usepackage{units}
\usepackage{xcolor}
\usepackage[caption=false]{subfig}
\usepackage[footnotesize,it]{caption}
\usepackage{paralist}
\usepackage{amsthm}
\usepackage{amsmath}
\usepackage{amsfonts}
\usepackage[numbers,sort]{natbib}
\usepackage{siunitx}
\usepackage[small,compact]{titlesec}
\usepackage{enumitem}
\usepackage{booktabs}
\setitemize{itemsep=1pt,topsep=1pt,parsep=1pt,partopsep=1pt}
\newcommand{\bi}{\begin{itemize}}
\newcommand{\ei}{\end{itemize}}

\newcommand{\eg}{{\it e.g.,}\xspace}
\newcommand{\ie}{{\it i.e.,}\xspace}
\newcommand\eat[1]{}

\newcommand{\systemname}{AutoTune\xspace}
\newcommand{\name}{\systemname}

\newcommand{\allnotes}[1]{}
\renewcommand{\allnotes}[1]{\textit{#1}}
\newcommand{\fixme}[1]{\allnotes{\bf\textcolor{red}{[FIXME: #1]}}}

\newcommand{\notepanda}[1]{\allnotes{\textcolor{cyan}{[Panda: #1]}}}
\newcommand{\notemichael}[1]{\allnotes{\textcolor{purple}{[Michael: #1]}}}
\usepackage{array}
\usepackage[printwatermark]{xwatermark}
\newcolumntype{P}[1]{>{\RaggedLeft\hspace{0pt}}p{#1}}
\usepackage{enumitem}
\setitemize{noitemsep,topsep=0pt,parsep=0pt,partopsep=0pt}
\hypersetup{draft}
\usepackage{authblk}
\makeatletter
\renewcommand\AB@affilsepx{, \protect\Affilfont}
\makeatother

\begin{document}
\title{AutoTune: Improving End-to-end Performance and Resource Efficiency for Microservice Applications}

\author[1]{Michael Alan Chang}
\author[2]{Aurojit Panda}
\author[3]{Hantao Wang}
\author[1]{Yuan Cheng Tsai}
\author[4]{Rahul Balakrishnan}
\author[1,5]{Scott Shenker}
\affil[1]{UC Berkeley}
\affil[2]{NYU Courant}
\affil[3]{Square}
\affil[4]{University of Illinois Urbana-Champaign}
\affil[5]{International Computer Science Institute}
\date{}                     
\setcounter{Maxaffil}{0}

\maketitle
\begin{abstract}
  Most large web-scale applications are now built by composing collections (from a few up to 100s or 1000s) of microservices. Operators need to decide how many resources are allocated to each microservice, and these allocations can have a large impact on application performance. Manually determining allocations that are both cost-efficient and meet performance requirements is challenging, even for experienced operators. In this paper we present \systemname, an end-to-end tool that automatically minimizes resource utilization while maintaining good application performance.
\end{abstract}
\section{Introduction}\label{sec:introduction}

\subsection{Motivation}
Most modern web-scale applications -- including those offered by Google~\cite{Burns2016BorgOA}, Uber~\cite{uber}, and Netflix~\cite{netflix} -- are comprised of multiple (and sometimes many) microservices, such as web servers, caches, load balancers, and data management systems. 
Operators rely on \emph{microservice orchestrators} such as Kubernetes~\cite{kubernetes} to connect and manage these microservices. Operators provide the orchestrator with an application specification -- listing the required set of microservices, their resource requirements, and any placement constraints -- and a cluster specification that describes the set of servers on which the application can be launched. Microservice orchestrators provide additional APIs that provide fine-grained control over an application's resource allocation and placement (which controls which components are colocated). 

In theory, operators can use these APIs (along with application APIs to change various configuration parameters) to achieve good application performance and resource efficiency.
However, this is not an easy task, because the end-to-end performance of microservice-based applications is an unknown and complex function of the resources (such as memory and cores) allocated to each microservice and, to a lesser degree, the microservice placement. There has been little progress in deriving analytical expressions for real-world application performance as a function of resources and placement. Faced with this unresolved complexity, many operators merely overprovision with resources~\cite{Dean2013TheTA, Burns2016BorgOA, Delimitrou:2014:QRQ:2541940.2541941, Jyothi2016MorpheusTA} and use simple affinity rules for placement~\cite{2019arXiv190903130S,redhat,kubernetes}. As we will show, this results in good performance but an inefficient use of resources. To rectify this, we have designed a tool called \systemname for \emph{automatically} reducing resource consumption while preserving adequate performance. 

\subsection{Considerations}

\systemname's design is strongly influenced by three basic considerations. First, we want a solution that can be used by non-expert operators, and is applicable across a wide variety of systems and use-cases. We therefore treat microservices as blackboxes, rather than assuming that detailed models of their performance are available. Similarly, we directly measure application performance rather than attempt to derive it from performance measurements on individual microservices. This allows the operator to select the performance objective most relevant to them (makespan, tail latency, etc.).

Second, we assume that operators can produce (based on previous experience running the application) an initial deployment (\ie a specification of the placement and the resource allocations for each microservice) of their application that achieves acceptable performance.  We take the performance of this initial deployment as the baseline performance expectation. Our goal is to find more efficient deployments that use fewer resources, yet achieves similar or better performance.

Third, in order to not interfere with the production environment, we intend \systemname to be run on a testbed where various configurations can be evaluated; such testbed experiments can easily be carried out in public clouds. In order for these testbed runs to be useful, we assume that the operator has access to a representative workload (or set of workloads) where achieving good performance and resource efficiency on these test workloads results in similar benefits in production use.  This is clearly true for recurring batch jobs, which have similar workloads across invocations. We later discuss how \systemname can apply to more varying workloads, which we then evaluate using a production trace.

\subsection{Approach}

Our goal in designing \systemname is not to achieve {\em provably optimal} performance or {\em provably optimal} efficiency, but to provide a widely-applicable and easy-to-use tool that can provide {\em as good or better} performance than the operator-provided baseline while typically reducing the number of servers required for deployment. Our approach to finding efficient deployments is comprised of two separate phases. 

In phase 1, which we call {\em resource clampdown}, we try to identify overprovisioned resources where we can reduce the resource dedicated to an individual microservice without impacting overall application performance. Once we have eliminated all overprovisioning, we then try to place microservices on the minimal number of servers while respecting these resource allocations. We then verify that the resulting deployment matches the performance of the initial deployment, and make adjustments if necessary to make this hold. 

In phase 2, which we call {\em performance improvement}, we take the deployment from phase 1 and test whether performance can be improved by assigning leftover resources and shifting resources on a server between microservices. The result is a locally optimal allocation of resources that equals or exceeds the original performance goal. 

We demonstrate the efficacy of our approach by using \systemname on three representative microservice applications exhibiting a variety of microservice design patterns with widely different microservices (both third-party and custom). We demonstrate that, for these use-cases, \name can reduce the number of servers by up to $6.6\times$, while simultaneously improving performance by up to $20\%$. 

\name's resource allocation and placement are computed for a specific workload, but in practice workloads change over time. In \S\ref{sec:workload} we demonstrate how \name can be used to overprovision resource so it can handle workload changes. We find that overprovisioning resources by $30\%$ is sufficient to handle a workload that has $1.6\times$. Additionally, we analyzed a a four-week long production trace from Datadog~\cite{datadog} and found that in that trace a workload increase of this magnitude occurs approximately every 51 hours. Thus, our analysis demonstrates that slight overprovisioning, and infrequently running \name is sufficient for handling dynamic workloads.

\vspace*{0.3in}

\section{Background}\label{sec:bg}

\eat{\notepanda{Changed, putting previous one in comments}
We begin by describing current approaches to automatically scaling microservice applications. These approaches addresses a different and complementary problem to ours, aiming to discover resource allocations that are sufficient for meeting a specified performance objective. We demonstrate that widely implemented autoscalers, which depend on local optimization, can allocate more resources than necessary to meet a given performance goal, while more global strategies, such as ones based on Bayesian optimization, cannot scale to handle applications with multiple microservices. This motivates the need for \name, which improves resource efficiency without negatively impacting performance.}

We begin by describing current approaches to automatically scaling microservice applications --  autoscaling and Bayesian optimization -- and observe that they will not find efficient deployments for many microservice applications. This lack of applicability motivates the need for \name. However, note that these approaches solve a subtly different problem than \systemname in that they typically start with some configuration which is underperforming and then add resources until the performance is adequate; \systemname starts with a configuration where the performance is adequate and then reduces (and rearranges) resources while maintaining (or improving upon) acceptable performance. The two approaches are complementary, and in \systemname when there is a sudden load spike we can fall back on autoscaling to cope.

\subsection{Autoscaling}\label{sec:bg:perf}
Some orchestrators (including Kubernetes) and most cloud providers (including Amazon~\cite{aws:elb}, Azure~\cite{azure:autoscale}, and Google~\cite{google:autoscale}) provide autoscaling services that observe each microservice's performance or utilization, and use this information to decide when to add additional instances of that microservice. These services accept as input a pre-defined threshold (\eg an acceptable response latency or utilization level) and increase the number of instances for a microservice whenever its performance drops below the threshold or its utilization exceeds the threshold.

Regardless of what metric is used to trigger scaling, existing horizontal and vertical autoscaling solutions focus on \emph{local optimizations}: they consider the performance or utilization of an individual microservice, without regard to overall application performance. Additionally, when triggered they scale only the \emph{triggering microservice}, without regard to how this might impact overall performance. This focus on local optimization has two shortcomings. 

First, using performance metrics requires the applications operator to decompose an application level performance requirement into performance targets for each individual microservice. As has previously been observed~\cite{cedar, jalaparti2013speeding}, the amount of time processing a request at different services depends not just on the application, but on various factors such as the request itself (which might dictate whether processing is done on the fast path or slow path) and the history of previous requests (which dictate whether a microservice might experience pauses due to garbage collection or other effects). Finding an appropriate decomposition for complicated real-world applications is thus quite challenging.

Second, the use of local optimization (whether performance or utilization driven) can result in suboptimal resource allocations, \ie in situations where an application is allocated more resources than required to meet a particular performance goal. We now illustrate this for utilization-triggered autoscaling.

\begin{figure*}[t]
\minipage{0.325\textwidth}
    \includegraphics[width=\linewidth]{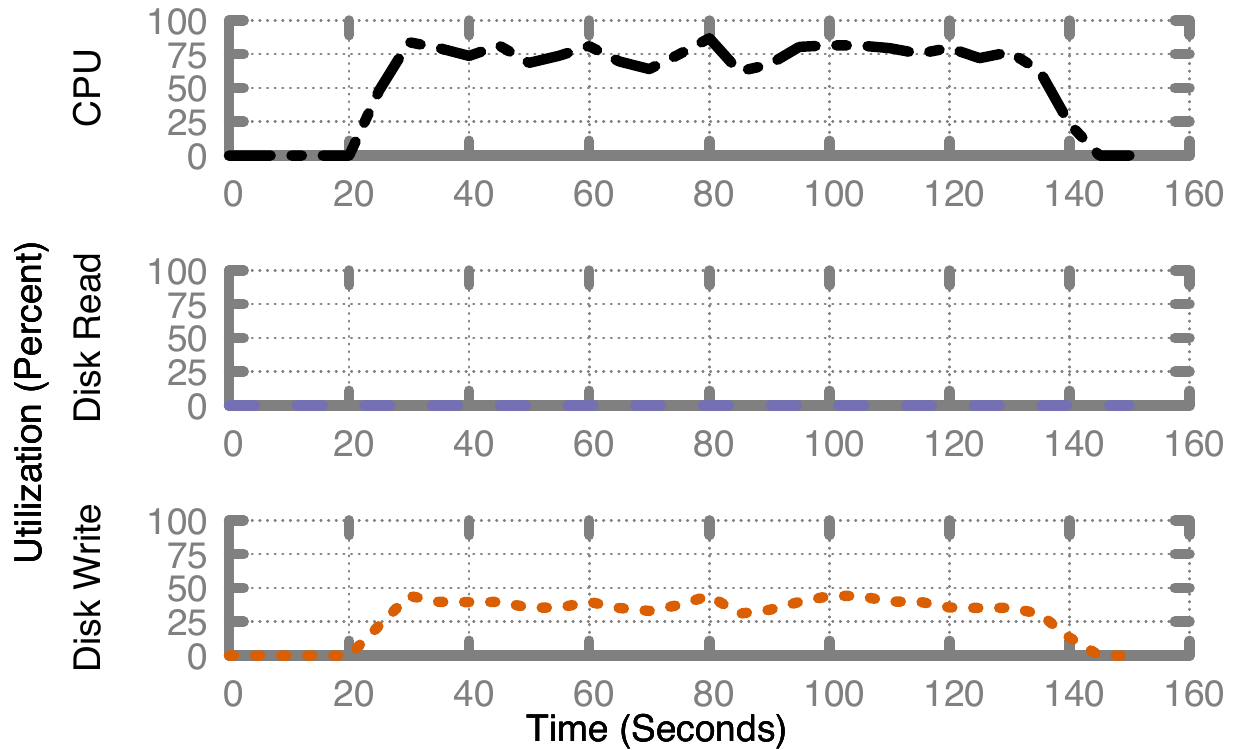}
    \caption{Observed utilization for frontend microservice.}\label{fig:util-flaskapp}
    \vspace{-0.1in}
\endminipage\hfill
\minipage{0.325\textwidth}
    \includegraphics[width=\linewidth]{plots/heterogenous-fioapp.pdf}
    \caption{Observed utilization for storage microservice}\label{fig:util-fioapp}
    \vspace{-0.1in}
\endminipage\hfill
\minipage{0.325\textwidth}
    \includegraphics[width=\linewidth]{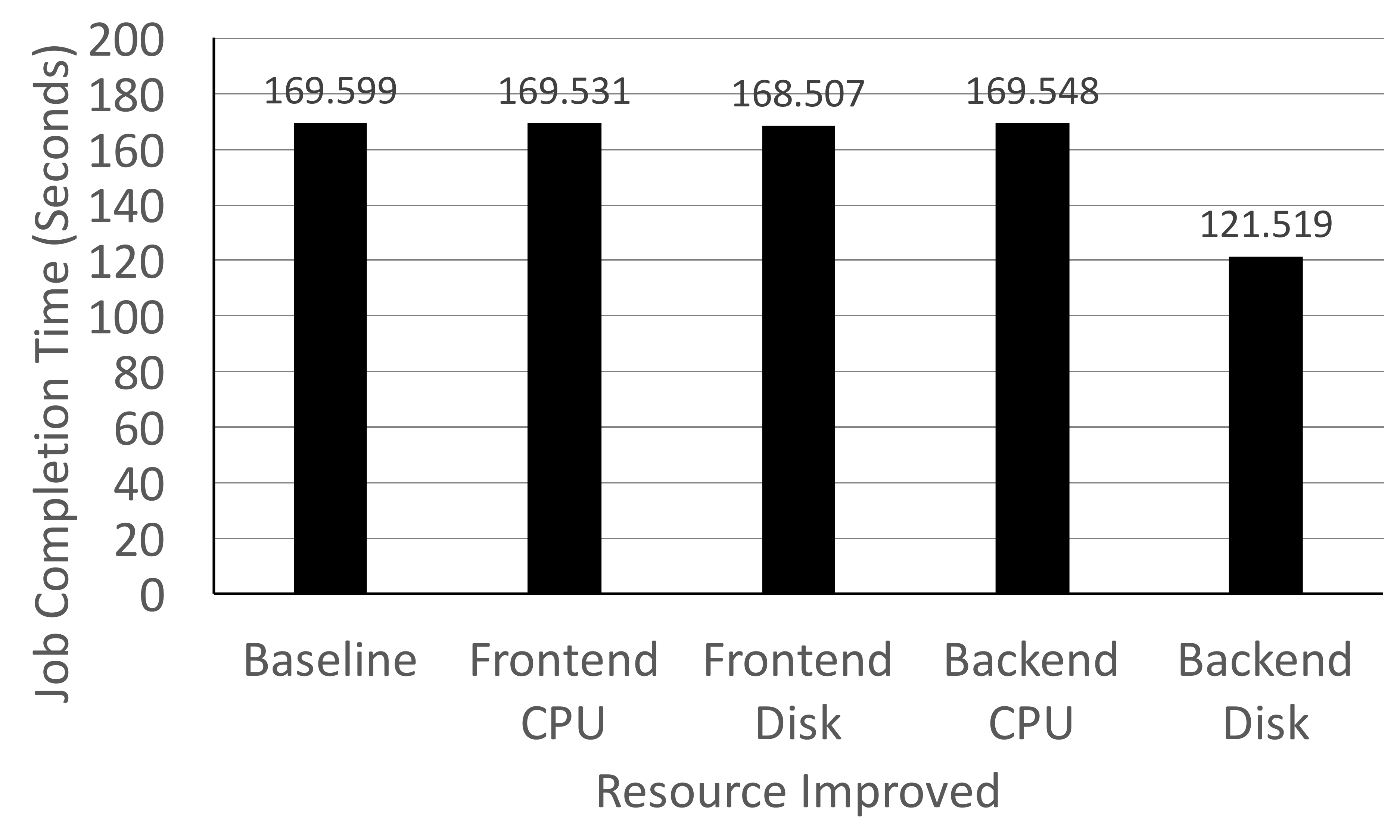}
    \caption{Actual Job Completion Time after adding resources to the multiservice microbenchmark.}\label{fig:hetero-improvement}
    \vspace{-0.1in}
\endminipage
    \vspace{-0.1in}
\end{figure*}

Consider an application comprised of two microservices: a frontend microservice and a storage microservice. The frontend microservice receives client requests and performs a blocking remote read from the storage microservice, which on receiving a read request, reads and processes the request file before returning the processed results. We collected resource utilization and job-completion times for each microservice during a baseline run, which we show in Figure~\ref{fig:util-flaskapp} and Figure~\ref{fig:util-fioapp}. Utilization seems to indicate that bottleneck resource is either (a) the CPU on the frontend microservice; (b) the CPU on the storage microservice; or (c) the \emph{disk} on the frontend microservice (used for logging). We then attempt to empirically validate these findings by increasing each of these resources and measuring their impact on the application's performance (Figure~\ref{fig:hetero-improvement}). We find, contrary to what the utilization measurements suggest, that none of them has any impact on application performance. In fact, the highest performance improvement comes from increasing \emph{disk bandwidth} on the storage microservice, a resource that was not highly utilized! This is because the frontend is blocked while waiting for the storage service to read data, and utilization is not sufficient to capture this dependency. 

While a microservice's performance (\eg a microservice's response latency) is a more relevant measure than its utilization, autoscaling based on such per-microservice performance metrics suffers from the same flaw of not capturing dependencies, and can thus result in inefficient provisioning.

\eat{Below, in \S\ref{sec:bg:utilization}, we discuss how this arises in  utilization triggered autoscaling; here we discuss performance triggered autoscaling.
Consider an application comprised of two microservices, a frontend webserver and a backend storage server. All user requests are first processed by the webserver,  which must then retrieve content from the backend storage server. We assume that both microservices scale linearly with additional resources. Finally, we vary the minimum request processing time for each microservice in order to evaluate different application configurations. Given this application we consider two performance-based autoscaling configurations: in the first autoscaling is triggered based on response latency. For the webserver this is the sum of the processing time at the webserver, and the storage server's response latency. In the second autoscaling is triggered based on the processing time at each microservice and is hence unaffected by the other microservice. In both cases, given an application level latency objective, we derive per-microservice latency thresholds by proportionately dividing the application level objective in the ratio of each microservices minimum response latency (or minimum processing time). In Figures \fixme{} we show the number of resources used in order to meet a range of latency targets, relative to what is needed when using \name (which focuses on global optimization). The figures show that \fixme{Fill in with figure...}}

\eat{Thus, in answer to the oft-asked question about utilization, we answer that there must be a tool that contextualizes the importance of a particular microservice and resource to overall application performance, and \systemname does to by measuring this performance directly.}

\eat{Operators often monitor the utilization levels (CPU, memory, disk, and network) in all the servers, and this can help them identify resources that might be performance bottlenecks. However, while such measurements are generally useful, high utilization levels do not necessarily indicate that a fully utilized resource is causing poor microservice performance. More generally, without an analytical model of application performance, one cannot hope that any set of  local measurements can fully capture the impact of a particular resource or microservice on the overall application performance. We discuss these observations in more detail in Section \ref{sec:utilization}.}

\subsection{Bayesian Optimization}

In recent work \cite{Alipourfard2017CherryPickAU}, Bayesian optimization techniques were used to improve the performance of an application consisting of a single service (Spark executors) with a single microservice per server. Optimizing this deployment merely requires determining resource allocations for that single microservice; this results in a relatively low-dimensional search space on which Bayesian optimization performs well. We seek a tool that can apply to applications with a heterogeneous set of microservices which interact with each other in a variety of ways. Such a setting results in a high-dimensional search space, making it impractical to apply Bayesian optimization. Furthermore, when dealing with a heterogeneous set of microservices, the mechanisms controlling scaling and placement decisions must consider resource and operational constraints, and these are difficult to incorporate in such Bayesian optimization algorithms. In contrast, \systemname uses a less sophisticated search algorithm than Bayesian optimization, but can handle a high-dimensional search space and complicated operational constraints.

\eat{The resource constraints in this case allows the search algorithm to ignore solutions where any server's resources  are overprovisioned, while operational constraints allow developers and operators to  constrain placement in  order to ensure independent failure domains, etc.
We ave not been considered when applying Bayesian optimization to this problem ( such as Cherrypick~\cite{Alipourfard2017CherryPickAU} which considered homogenous distributed applications such as data processing frameworks. We attempted to extend these Bayesian approaches to such heterogeneous applications, we failed to find a good mechanism for applying these constraints, and instead found that Bayesian Optimization spent a large portion of its search time ruling out infeasible placements due to its inability to consider such constraints. Our approach on the other hand uses a less sophisticated search algorithm, but provides a natural mechanism for imposing such constraints which in turn drastically reduces the search space, which in turn enables both faster convergence to a final configuration and wider applicability.}


\section{The Design of \systemname}
\label{sec:design}
Having reviewed existing approaches to optimizing application performance and resource efficiency, we now describe \name's design. We begin by defining terminology used through the rest of the paper, then describe the types of deployments we target, and finally present \name's design.

\subsection{Terminology}
\name's inputs are the application $A$ and performance metric $P$ to be optimized, an initial deployment $D_0$, and a representative workload $W$. The initial deployment consists of a resource allocation $R_0$ that specifies what resources are allocated to each microservice in $A$, and a set of placement decisions that specifies what microservices are colocated with each other. Given these inputs, \name uses $W$ to compute an initial performance $P_0$, and \name's optimization process ensures that any deployments perform at least as well or better than $P_0$.

\name considers four resources: memory, CPU cores (both full cores and fractional assignments), disk bandwidth, and network bandwidth. We use the term microservice resource (MR) to refer to the allocation of one of these resources to a particular microservice on a particular server. \systemname adjusts microservice resource allocations and microservice placements in order to minimize the number of servers required and then produces a new deployment at the end of this process.

\subsection{Target Deployments}
We envision that \name is deployed in a test cluster, and the resulting deployment is then used in production. \name's computation relies on changing resource allocations, and observing the performance impact of such changes. Running this in a test cluster ensures that \name does not negatively impact performance in production. Additionally, \name is designed to be used in clusters where resources are allocated in the form of servers or VMs (we refer to both as servers below), and the orchestrator can place zero or more microservices in each VM. As a result, we measure resource efficiency in terms of servers, and \name's optimization objective aims to minimize the number of servers, rather than merely minimizing total allocated resources. 

\subsection{Phase 1: Resource Clampdown}
\label{sec:clampdown}
\name's first phase focuses on identifying over-provisioned microservice resources in the initial deployment, and then eliminating over-provisioning. 

\name identifies over-provisioned resource by \emph{stressing} individual MRs; \ie reducing the amount of the resource allocated to the microservice and then measuring the end-to-end application performance. \name uses the results from stressing to classify each MR as either an {\em impacted microservice resource} (IMR) -- a MR whose reduction hurts application performance -- or a {\em non-impacted microservice resources} (NIMR) -- a MR whose reduction has no negative impact on application performance. Whenever a NIMR is identified, \name reduces the resource allocation by the same amount used in the stressing step. \name repeats this process until no more NIMRs are identified. The resulting resource allocations are {\em tight}, in that any resource reduction will result in performance degradation. 

While the previous step yields a tight resource allocation, it does not necessarily free up any servers, since it does not change placement. The next part of phase 1 thus produces a new placement that packs all of the microservices into fewer servers. While doing so, \name attempts to avoid colocating any pair of microservices that might affect each other's performance. In order to do so we rely on the ability to order the impact of different microservice resources. We define a microsevice resource $r$ as being more impacted than MR $r'$ if reducing $r$'s allocation has a greater impact on application performance than $r'$. Given this ordering, we can also identify the set of most impacted MR's. 

Observe that the stressing results above already provide us with sufficient information to identify the most impacted resource for each microservice. Given this information, and the number of available servers, the \name placement algorithm begins by placing microservices so that any two colocated microservices differ in their most impacted MR. When this placement strategy is no longer feasible, \name switches to a round-robin placement strategy. The resulting placement is guaranteed to fit within only as many servers as are needed for the tight resource allocation identified previously; however, this change in placement might have resulted in worse performance.

To protect against this, \name checks application performance with the computed placement. If the new performance is worse than the initial performance $P_0$, then \name tries other variants (by changing the order in which microservices are placed) to identify a placement that performs as well or better than $P_0$. If no such placement can be found, \name increases the number of servers and reruns the placement strategy. This process necessarily terminates, since the initial placement had performance $P_0$. 

At the end of phase 1, \name produces a new deployment where no MR is overprovisioned, and that potentially uses fewer servers without negatively impacting application performance.

\subsection{Phase 2: Performance Improvement}
\label{sec:climbing}

The deployment produced in phase 1 might not utilize all server resources. In the second phase \name allocates these resources so as to improve application performance. This stage begins with \name assigning unallocated resources on each server to the microservice most likely to benefit from additional resources. To do so, for each type of resource available on a server, \name identifies the microservice on that server that is most impacted by that resource type, and allocates the microservice all available resources of that type. \name reuses measurements from phase 1 to determine impacted microservices on a server.

While the previous step allocates all available resources, in many cases performance can be further improved by changing resource allocations. This is because in cases where two microservices $\alpha$ and $\beta$ on the same server are impacted by the same resource, then it is possible that moving impacted resources from one service (\eg $\beta$) to the other (\eg $\alpha$) results in a net improvement in application performance. We identify opportunities for transferring resources using a gradient descent algorithm which iteratively matches highly sensitive MRs with less sensitive MRs of the same resource type that are colocated on the same machine.

\eat{
\begin{lstlisting}[mathescape=true,caption={\notemichael{Propose to cut this algorithm from the paper for space} Algorithm for Gradient Descent from Performance Improvement Phase. The algorithm starts from the placement and resource allocation determined in Phase 1, and then adjusts these allocations based on transfers. The function Gradient(R) produces a vector of changes in performance based on discrete changes in each MR (where the step size is $\delta_r$ for resource $r$), and the sort function orders them in decreasing order.}, label=lst:resource_constrained_alg, basicstyle=\small,float,escapeinside={(*}{*)},belowskip=-0.8\baselineskip]
Given: 
  Starting Resource Allocation: (* $R$ *)
  Performance Function: (* $P$ *) 
  Step size: $\delta_r$ for resource $r$

do:
  PrevP = (* $P(R)$ *)
  max_impact_list = sort(Gradient(R))
  min_impact_list = reverse(max_impact_list)
  transfer_found = False
  for m_h in max_impact_list:
     for m_l in min_impact_list with m_l < m_h:
        if MRs are colocated, 
           and have same resource type $r$:
            R[max_IMR] += $\delta_r$
            R[min_IMR] -= $\delta_r$
            transfer_found = True
            break
     if transfer_found:
        break
  CurrentP = (* $P(R)$ *)
while (CurrentP - PrevP > threshold)
\end{lstlisting}}

\section{Additional Issues}
\label{sec:additional}
Next we discuss some of the additional challenges addressed by our design.

\subsection{Reducing Search Space}
Because we have four resources, stressing microservice resources one at a time would require that \name evaluate performance for $4M$ different deployments (where $M$ is the number of microservices). \name reduces the number of deployments that have to be tested by randomly partitioning all MRs into $p$ partitions. \name then measures performance after reducing allocation for {\em all MRs} in a partition. This measured performance is used differently by the two phases:

\noindent\textbf{Phase 1: Partitions in Resource Clampdown} If application performance does not degrade when a partition is stressed, \name infers that none of the MRs in the partition are impacted. If on the other hand a performance degrades, \name splits the partition itself into $p$ random partitions, and uses the same process to select those subpartitions which degrade performance. This process continues recursively until all impacted MRs are found.

\noindent\textbf{Phase 2: Partitions in Performance Improvement} When using gradient descent, \systemname first identifies the most-impacted partition. Within that partition, \systemname applies the original gradient descent method, stressing each MR individually. Thus, in each iteration of the the gradient descent process, \systemname with the pruning method with $p$ partitions only needs to explore $\frac{N \cdot k}{p} + p$ MRs per iteration, rather than of having to explore $N \cdot k$ MRs. 

Observe that for partitioning to be effective in either phase we need most partitions to not impact application performance. In our tests we have found that empirically this is indeed the case, and the set of impacted MRs is relatively small. In our experiments we found that the use of partitioning reduced \name's runtime by $2-3\times$ on average. However, even when there are many impacted MRs the use of partitioning has no impact on correctness, nor does it significantly impact \name's efficiency. 


\subsection{Minimizing Placement Changes}
Changing the placement of a microservice takes longer than merely changing its resource allocation. This is because changes to placement necessitate launching a new copy of the microservice, while resource changes can often be done without needing to terminate an existing copy. This observation influences our design in two ways: first, we minimize the number of times placement is changed (only at the end of phase 1 and the beginning of phase 2); second, we stress (\ie reduce allocated resources) rather than increase resources to measure a MR's impact on application performance. Stressing a resource is guaranteed not require placement changes, while increasing resource allocation might in cases where a microservice is running on a fully utilized server.

\subsection{Discrete Gradient Descent}

Analytically, gradient descent uses derivatives in order to choose the direction for change. Here, we must implement a discrete derivative (when stressing the application); in order to evaluate the derivative, we compare the application performance with the original MR with one that differs by an amount $\delta$, and the question is how do we choose $\delta$. \systemname sets $\delta$ proportional to the server's resource capacity. However, rather than use a fixed percentage, we start at 30\% and then, at each iteration of the algorithm, we decrease this percentage by a factor of 1.2. This allows us to take finer and finer tests of discrete changes. We also implement a binary search feature when evaluating resource transfers. When we increase one MR by $\delta$ and decrease another by the same amount and find no change in application performance, we then execute a binary search between the original and tested allocations to see if we overstepped a local optima.

\subsection{Placement}
 Our experience suggests that placement is typically not a crucial factor in application performance, aside from the special case of affinity. However, because containers (and indeed most other software isolation mechanisms) do not completely isolate performance, interference between colocated microservices can negatively impact application performance. Discovering this would require exhaustively searching through all placement possibilities, which would be too time-consuming. Instead, we address this issue by preferring, when when searching for optimized deployments, those that colocate microservices that were colocated in the initial deployment and enforcing any administrator provided placement constraints. This allows \systemname to incorporate any insights the operator might have about possible performance interference.

\section{Dynamic Workloads}
\label{sec:workload}
Thus far our design has focused on applications with relatively predictable load patterns where we could assume that the operator supplied a representative workload which could be used to evaluate performance. Since almost all workloads have some natural variation, we further assume that this supplied workload represents the high-end of anticipated workloads (and the degree of such overestimation is up to the operator based on their tradeoff between performance and efficiency). We also assumed that the operator provided an initial deployment on which this workload has acceptable performance. These are reasonable assumptions for batch workloads and relatively stable real-time workloads. However, we cannot make these assumptions for real-time workloads with significant variation (which we assume is the common case for real-time workloads).

For these real-time workloads, the common practice is to significantly overprovision (we have heard some operators provision their deployment to handle up to three times the average workload), and then rely on autoscaling (as described in Section \ref{sec:bg}, this independently scales the microservices based on local measurements) to handle any further overloading. We propose an alternate approach, based on \systemname, to increase efficiency. In this case, the operator provides several different workloads (based on historical usage patterns) and their corresponding deployments which achieve satisfactory performance. \systemname is run on each of these workload/deployment pairs, producing more efficient deployments which achieve the same or better performance than the given deployment for each workload. In production, the operator uses an automated mechanism that measures the current load and picks the most appropriate deployment for the current load. 

For this approach to be feasible, we must ensure that the deployment changes are relatively rare, which requires a degree of overprovisioning and using some amount of hysteresis so that the deployment is not changed based on short-term fluctuations. To investigate how this might work in practice, we study a production Datadog trace that contains the rate of API calls in 10 second intervals over a four week time period. 

We apply this trace to a web application developed using the MEAN Stack (described in \S\ref{sec:application}), and implement the following scheme. We pick some degree of overprovisioning (\ie pick a deployment designed for a load level higher than the average by some amount) and stick with the degree of overprovisioning until we observe five minutes of load that are significantly higher or lower than the previously observed average. The five minute period is unlikely to cause major performance disruptions because, as we show in \S\ref{sec:evaluation}, \systemname-generated deployments hold performance even when the workload is 10-20\% greater than experimental workload, and then performance degrades gracefully that.

When the aberrant load is sustained for more than five minutes, the system switches to a deployment that can handle the new load. This new deployment is similarly overprovisioned, so it can handle increased load in the future. Since we assume that \name only produces a finite number of deployment configurations, we pick the deployment generated using the workload with the lowest average request rate that is still higher (by the same overprovisioning factor) than the production workload's recently observed average request rate.

Table \ref{tab:resource_overprovisioning} shows how often the deployment would need to change, for different levels of over-estimating workload. If the operator overprovisions by  10\%, the system would have to transition to a new deployment once every 45 minutes. At the other extreme, if the operator overprovisions by a factor of three, the system would only have to transition to a new deployment once every week. 

Where in between these two extremes an operator would choose to operate would depend in part on how expensive (in terms of resources) it is to overprovision for higher loads. These results are also shown in Table \ref{tab:resource_overprovisioning}. Note that overprovisioning for a factor of three costs roughly 54.5\% extra in resources, and doing so for a factor of 2 costs roughly 34\% (and requires deployment changes roughly once every three days). 

\begin{figure}[t]
     \centering
     \includegraphics[width=0.4\textwidth]{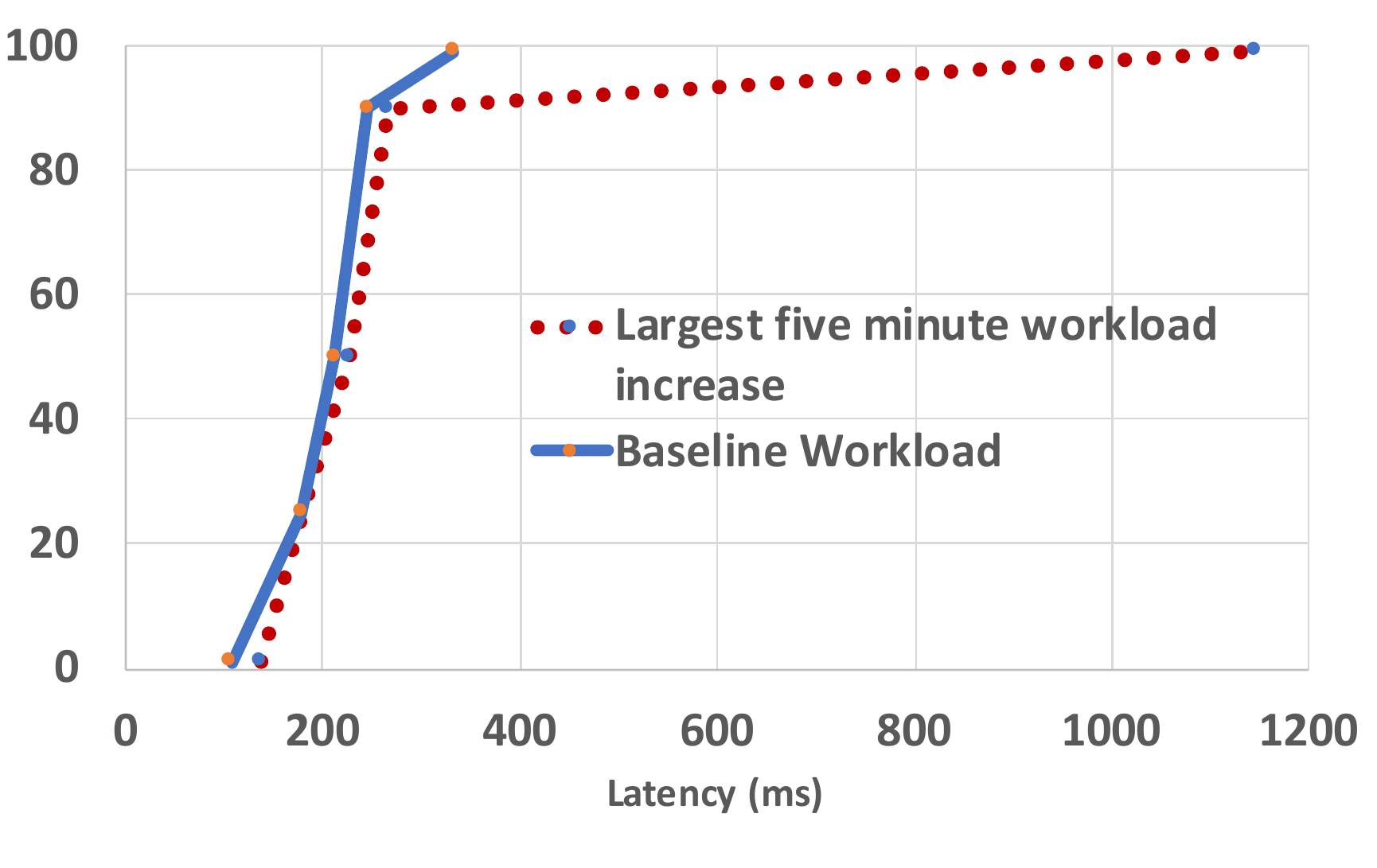}
     \caption{CDF of response latency for a 30\% overprovisioned deployment with the expected request rate (baseline) and a request rate that is $1.6\times$ higher (Largest five minute increase).}
     \label{fig:traceworkload}
     \vspace{-.2in}
 \end{figure}

Finally, we measure the impact of response latency to changes in workload in an overprovisioned setting, since too large an increase in response latency would pose an impediment to the use of \name in dynamic settings. For this evaluation we overprovisioned the cluster with 30\% additional resources, and then considered a scenario where request rates instantaneously increase to $1.6\times$ the original request rate. The $1.6\times$ increase represents the largest consistent increase (\ie sustained for 5 minutes) we observed in the Datadog trace. The request rates and distributions were taken from the trace. We show a CDF of response latencies for the original request rate (baseline) and the increased request rate in Figure~\ref{fig:traceworkload}. We observe that median and $90^{th}$ percentile response latencies remain unchanged. While we do observe a nearly $6\times$ increase in maximum latency, this is likely due to the additional request queuing. We can thus conclude that most requests see reasonable response latencies, and hypothesize that additional overprovisioning might be necessary in scenarios where low tail latencies are crucial.

Thus, assuming a production deployment similar to the one captured by the Datadog trace, one can use \systemname for dynamic workloads without significant overprovisioning and infrequent deployment changes. 

\begin{table}[th]
\footnotesize
\centering
\resizebox{\columnwidth}{!}{
\begin{tabular}{@{}rrrrrr@{}}
\toprule
\multicolumn{1}{c}{\textbf{Workload Increase}} & \multicolumn{1}{c}{\textbf{Percent Overprovisioned}} &
\multicolumn{1}{c}{\textbf{Time between change}}
\\ \midrule
$1.2\times$ & 0.01 & 1.49 hours \\
$1.4\times$ & 27.5 & 1.95 hours\\
$1.6\times$ & 29.86 & 51.3 hours\\
$1.8\times$ & 30.2 & 59.62 hours\\
$2\times$ & 34.0 & 77.88 hours\\
$3\times$ & 54.5 & 166.1 hours\\
\end{tabular}}
\caption{The impact of changes in workload: by how much do resources need to be overprovisioned to handle workload increases; and, on average, how much time (measured in hours) elapses in our trace before overprovisioned resources are no longer adequate.}
\label{tab:resource_overprovisioning}
\vspace{-0.2in}
\end{table}

\eat{However, we can also generalize \systemname to more dynamic workloads, where the variation is too great to account for with the above approach. We require that the production version of the application have real-time monitoring \textit{application performance} and \textit{per-microservice traffic volume}. We consider two cases where the production performance may degrade. \\
\eat{
\textbf{Constant Performance, Increased Traffic Volume}: This suggests that the application remains in a locally optimal regime even as traffic load increases. In section \S\ref{sec:evaluation}, we demonstrate that resource allocations determined by \systemname continue to provide comparable performance despite small workload fluctuations.\\ 
\textbf{Constant Performance, Decreased Traffic Volume}: In this case, the \systemname resource allocations might be overprovisioned. If the operator gauges that the decrease in traffic volume is both meaningful and sustained, she may opt to re-run \systemname with this decreased workload.\\}
\textbf{Worse Performance, Increased Traffic Volume}: In this case, we do not change the allocations of the existing containers (since this would require redeploying some of them); instead \systemname activates horizontal scaling (commonly deployed with Kubernetes\cite{k8s:ha}) to deploy more instances. During this transition, application performance is unstable and may degrade; this is an orthogonal problem that applies to horizonatal autoscaling under increased load.\\
\textbf{Worse Performance, Constant or Decreased Traffic Volume}: Here there may be external factors influencing performance, or the workload might have changed in ways not visible to the monitoring (\eg the nature of the queries may have changed to be more compute intensive). \\
\textbf{Operator Response: } For both cases, we note that these require no operator intervention during workload changes. The operator retains characteristics of the production traffic through approaches described in \S\ref{sec:implementation}, and can later elect to re-run \systemname in an offline fashion.}

\section{Implementation}
\label{sec:implementation}

Next we describe \name's prototype implementation. \name is designed so it can be integrated with any container orchestrator. Our prototype implementation integrates with two orchestrators: Kuberentes and Kelda~\cite{kelda}. Below we detail some of the interesting aspects of our prototype implementation.


\subsection{Workload Generation}
\label{subsec: workload_gen}
Similar to prior approaches for resource allocation~\cite{Alipourfard2017CherryPickAU}, \name requires that the operator provide a representative workload, which it then uses for optimization. In our evaluation (\S\ref{sec:evaluation}) we use Apache Bench~\cite{apachebench} as a workload generator, and this approach might be usable by many applications. In addition, operators can also use tools such as Kraken~\cite{Veeraraghavan2016KrakenLL}, GoReplay~\cite{goreplay}, etc. to record and replay production traffic when using \name.

Additionally, many organization rely on canary deployments for testing and profiling. Live production traffic is often mirrored to these canary deployments, thus allowing profiling and debugging on live traffic. \systemname can be used in canary environments. While this might increase the time for convergence due to traffic changes between experiments, this does not impact correctness. Moreover, recent work ~\cite{10.1145/3126908.3126969} has shown that one can use transfer learning to extend performance models learned in more constrained environments (\eg in Canary deployments) to more general environment (\eg production clusters). In future work we plan to investigate the use of similar techniques in order to improve \name's robustness.

\subsection{Reconfiguring Applications}\label{sec:impl:reconfig}
The configuration for some microservies, \eg Nginx or Spark, depend on the number of resources allocated to them, and thus must be changed by \name. For such microservices, we require that operators provide scripts that \name can invoke to update microservice configuration whenever resource allocations are changed. Programmatic APIs to change configuration are increasingly common, andother tools including  Kubernetes Helm \cite{helm} rely on similar mechanisms.

\subsection{Stressing}\label{sec:imple:stress}
As we described previously in \S\ref{sec:design}, \name stresses microservice resources in order to identify impacted MRs. Our current implementation considers four resources: CPU core allocation, CPU quota per core, network bandwidth, disk bandwidth, and memory allocation. \systemname can be easily extended to consider other resources. Below we briefly describe how we stress each of these resources:

\paragraph*{CPU Quotas:} \name relies on \texttt{cgroups} to control CPU allocations, and allocates microservices both timeslices on a single core and whole cores. \name relies on Linux's completely fair scheduler (CFS) to allocate timeslices on a core by specififying a microservice's container period and quota. The current implementation allocates a quota relative to a fixed period (chosen at the start of the experiment) such that the maximal stressing never exceeds the minimal scheduling time quantum in the underlying operating system. 

\paragraph*{CPU Cores: } \name allocates CPU cores to a microservice by increasing its CPU quota sufficiently so that no other microservice can be colocated. More precisely, given a microservice with $quota$ aggregate CPU quota and $c$ cores, throttling down by one core would mean provisioning a quota of $\frac{quota}{c} \cdot (c-1)$.

\paragraph*{Memory stressing:} \systemname uses \texttt{cgroup}'s memory quotas to limit the amount of physical memory and swap space allocated to a microservice. In order to prevent out-of-memory errors, \name monitors each microservice's memory utilization and ensures that the application has sufficient memory and swap space allocated.


\paragraph*{Disk Bandwidth:} \systemname stresses the disk using \texttt{blkio}~\cite{redhat:cgroups}, and sets hard limits on read and write bandwidth from block devices. \systemname uses both joint and individual throttling of read and write bandwidth. 

\paragraph*{Network Bandwidth:} \systemname stresses the network by limiting link bandwidth Linux \texttt{tc}\cite{tc_man}. To do this we first measure the maximum attainable inter-VM network bandwidth, and then impose $k$\% stress by limiting the container's bandwidth to $(100-k)$\% of this maximum. \texttt{tc} uses hierarchical token bucket (HTB) to implement this rate limit, and scheduling network traffic using HTB imposes some CPU overhead. In our experience this additional overhead did not noticeably affect our results.

\begin{figure}[t]
    \centering
    \includegraphics[width=0.42\textwidth]{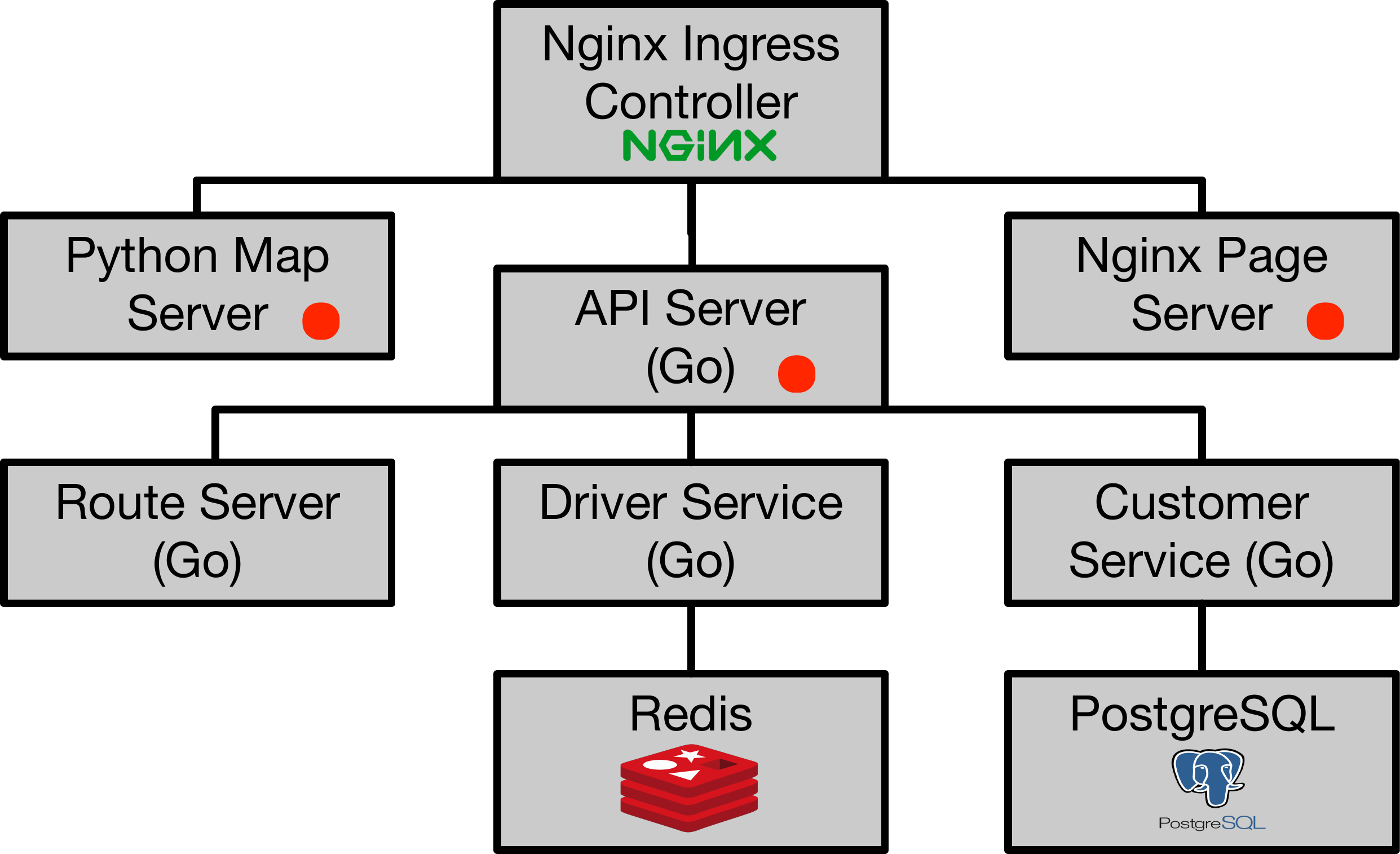}
    \caption{\textit{HotROD} Application, modified from Uber. Red circles indicate endpoints.}\label{fig:hotrod-app}
    \vspace{-.2in}
\end{figure}

\begin{figure}[t]
    \centering
    \includegraphics[width=0.39\textwidth]{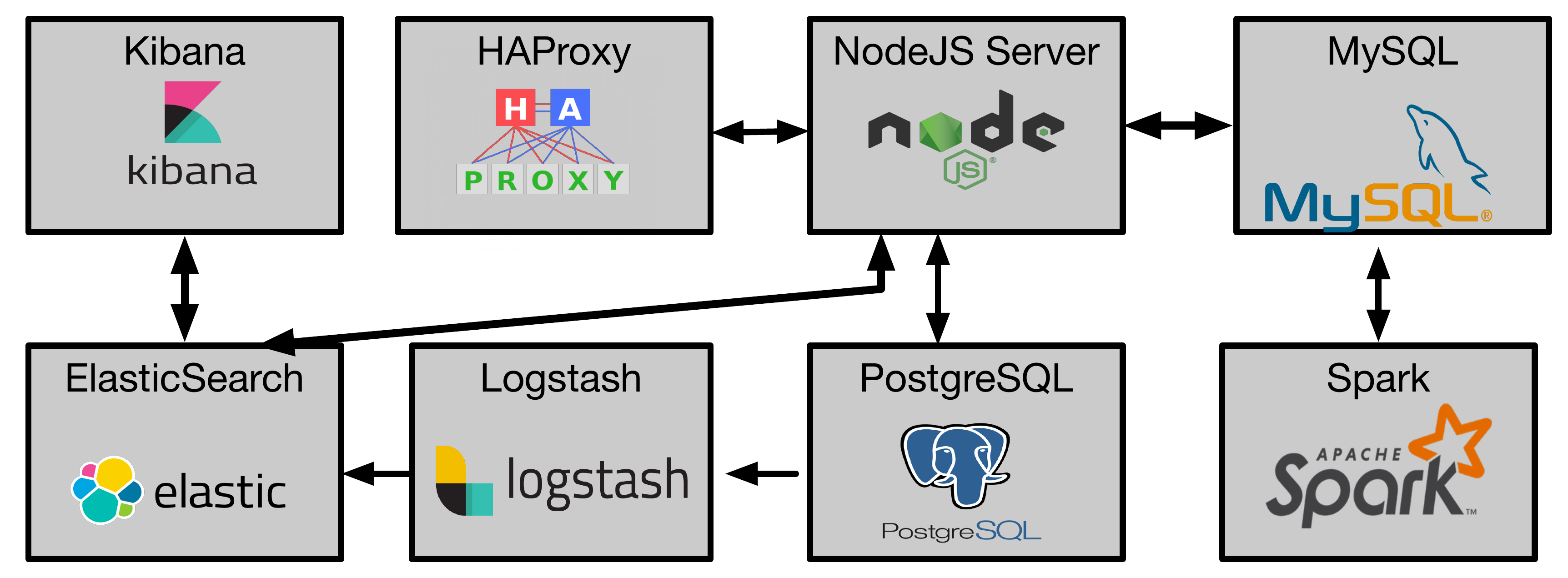}
    \caption{Apartment Rental Multi-service Application}\label{fig:apt-app}
    \vspace{-.2in}
\end{figure}

\section{Microservice Application Overview}
\label{sec:application}
For our evaluation we use microservice applications that resemble those running in production, and used by enterprises and moderate sized deployments, rather than services such as S3 and Google.
We first describe some common microservice design patterns that we observed, and then we present three end-to-end microservice applications that we used to evaluate \name.

We surveyed a number of industry practitioners to determine (i) which open source microservices were commonly used together, e.g., ELK (Elasticsearch, Logstash, Kibana) stack, Spark/MySQL, etc. and (ii) how these services are architected, e.g., chain or aggregation patterns \cite{microservicedesignpattern} \cite{microservicedesignpattern2}. In response, we selected three applications: two (slightly modified) open source microservice applications and one of our own.

\noindent\textbf{HotROD Application ~\cite{hotrodapp} (30 MR)} (Figure \ref{fig:hotrod-app}) is a modified distributed tracing application created by Uber to demonstrate its open source tool Jaegertracing \cite{hotrodapp}. We extended HotROD by adding two additional features: First, we split the frontend to be a separate API service and Nginx service to serve static pages. We also added a mapping service to pull graph data from S3 which is plotted using Networkx~\cite{networkx} and returns a JSON file which is rendered on the webpage. We also add an Nginx ingress and Haproxy load balancers to distribute load across all replicas of the mapping and API service. These extensions increase the number of MRs \name need consider when applied to HotROD.

\noindent\textbf{MEAN Stack ~\cite{todo-app} (12 MR)} is an application commonly used in tutorials introducing programmers to the MEAN stack, a popular server-side Javascript stack. The application uses three different types of microservices:  MongoDB (database service), NodeJs (web server), and HAProxy (load balancer). 

\noindent\textbf{Apartment Rental Application (48 MR)} (Figure~\ref{fig:apt-app}) is a web-application we developed to test this project, that is comprised of eight microservices. We use two different workload when testing this application: a write heavy workload (referred to as \textit{Apartment App Write}) and a workload consisting of a mix of reads and writes (referred to as \textit{Apartment App Mix}).


\section{Evaluation}
\label{sec:evaluation}
\begin{outline}
We evaluated \name using the applications described in \S\ref{sec:application}, and running them on clusters of \textit{m4.xlarge} AWS EC2 instances, each of which ran Ubuntu 17.04. Unless otherwise specified, we use $99^{th}$ percentile response latency as our performance metric. We have tested on a broader range of metrics, and observed similar results. Below we first present end-to-end evaluation results showing \name's efficacy, and then present microbenchmarks.


\begin{figure}[h]
\centering
\subfloat[MEAN Stack]{
	\label{fig:mean-e2e}
	\includegraphics[width=0.9\columnwidth]{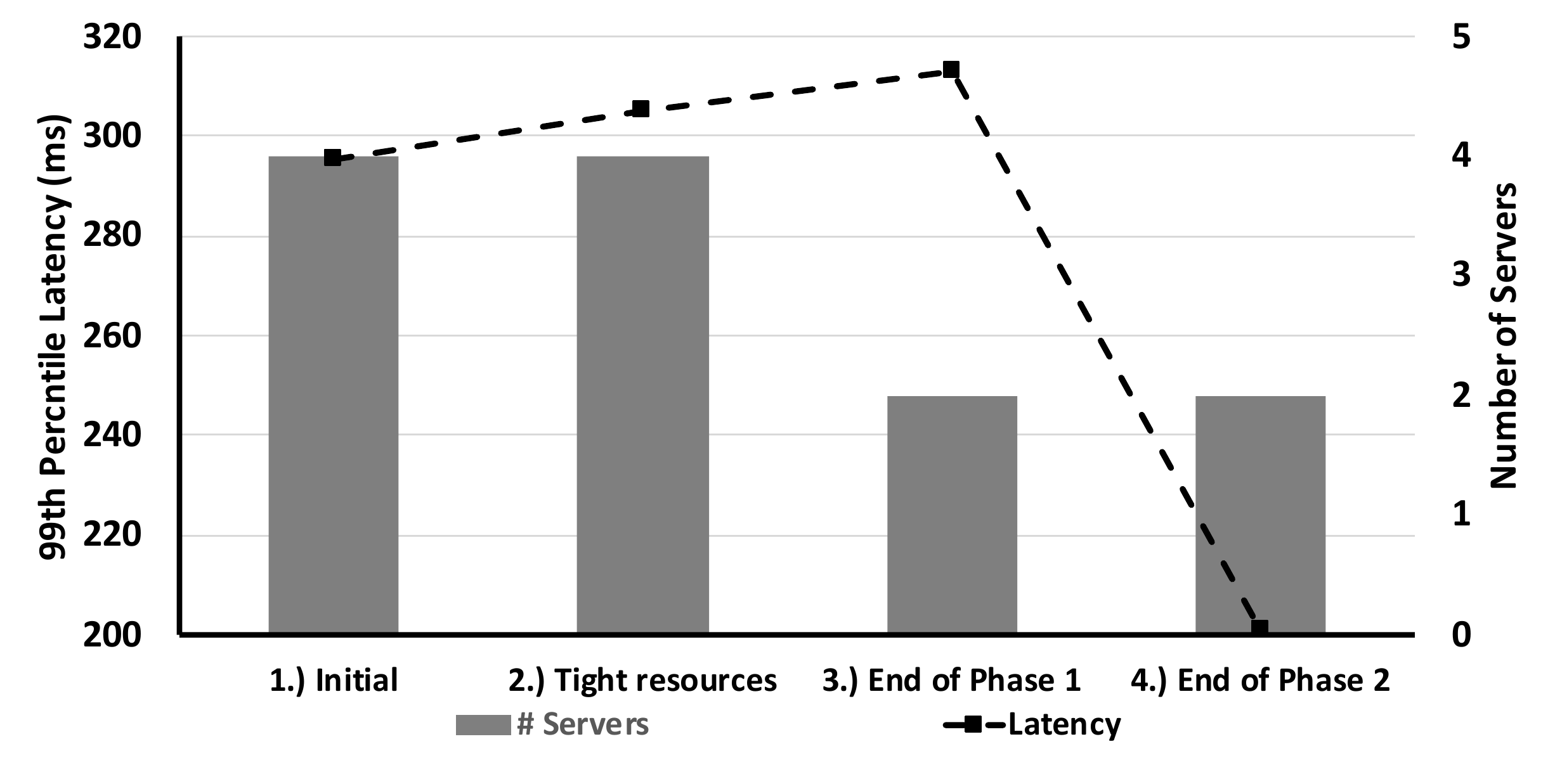} } 
    \vspace{-0.1in}
\subfloat[Apartment Application]{
	\label{fig:aptv2-e2e}
	\includegraphics[width=0.9\columnwidth]{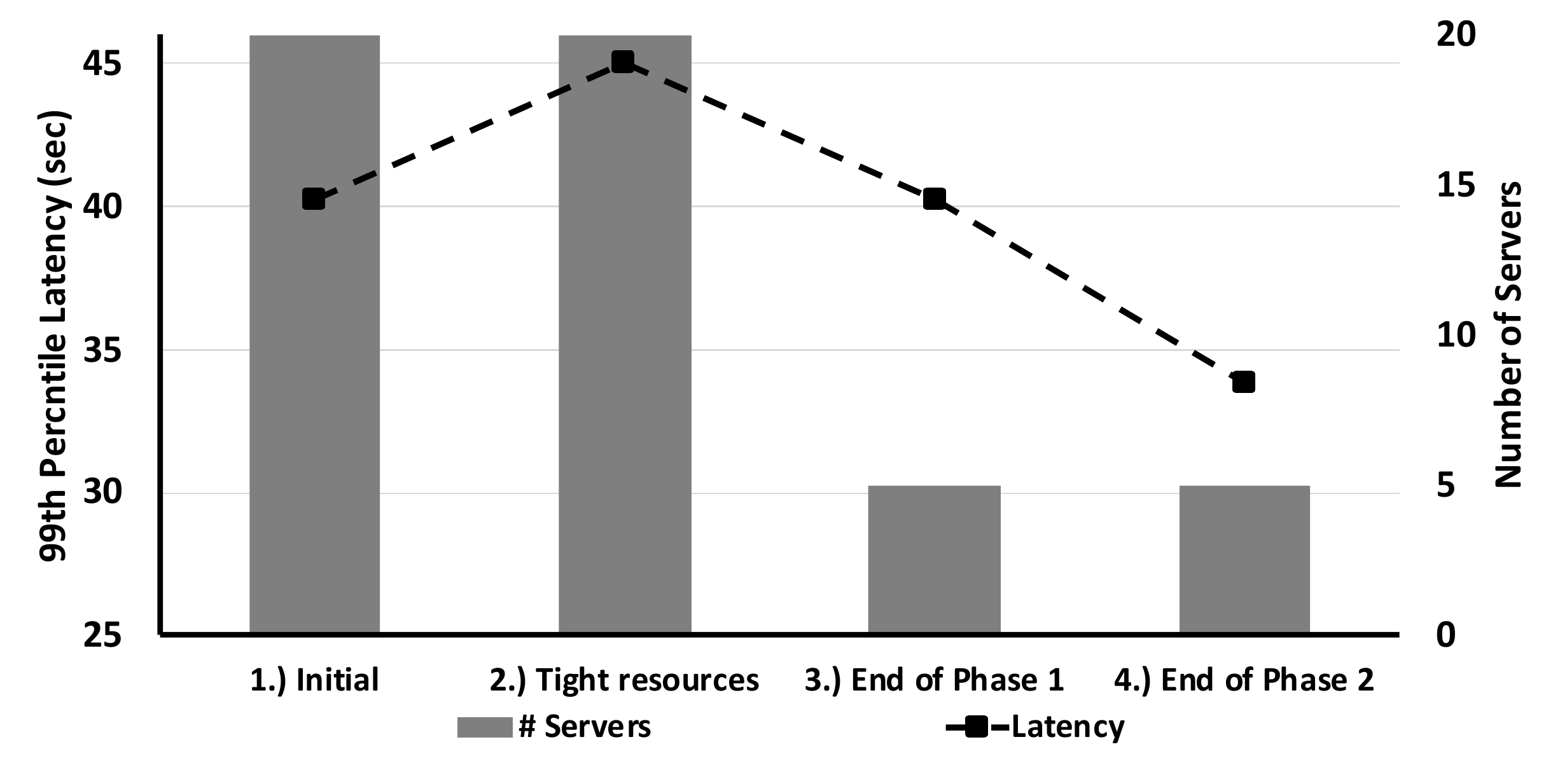} } 
    \vspace{-0.1in}
\subfloat[HotRod Application]{
	\label{fig:hotrod-e2e}
	\includegraphics[width=0.9\columnwidth]{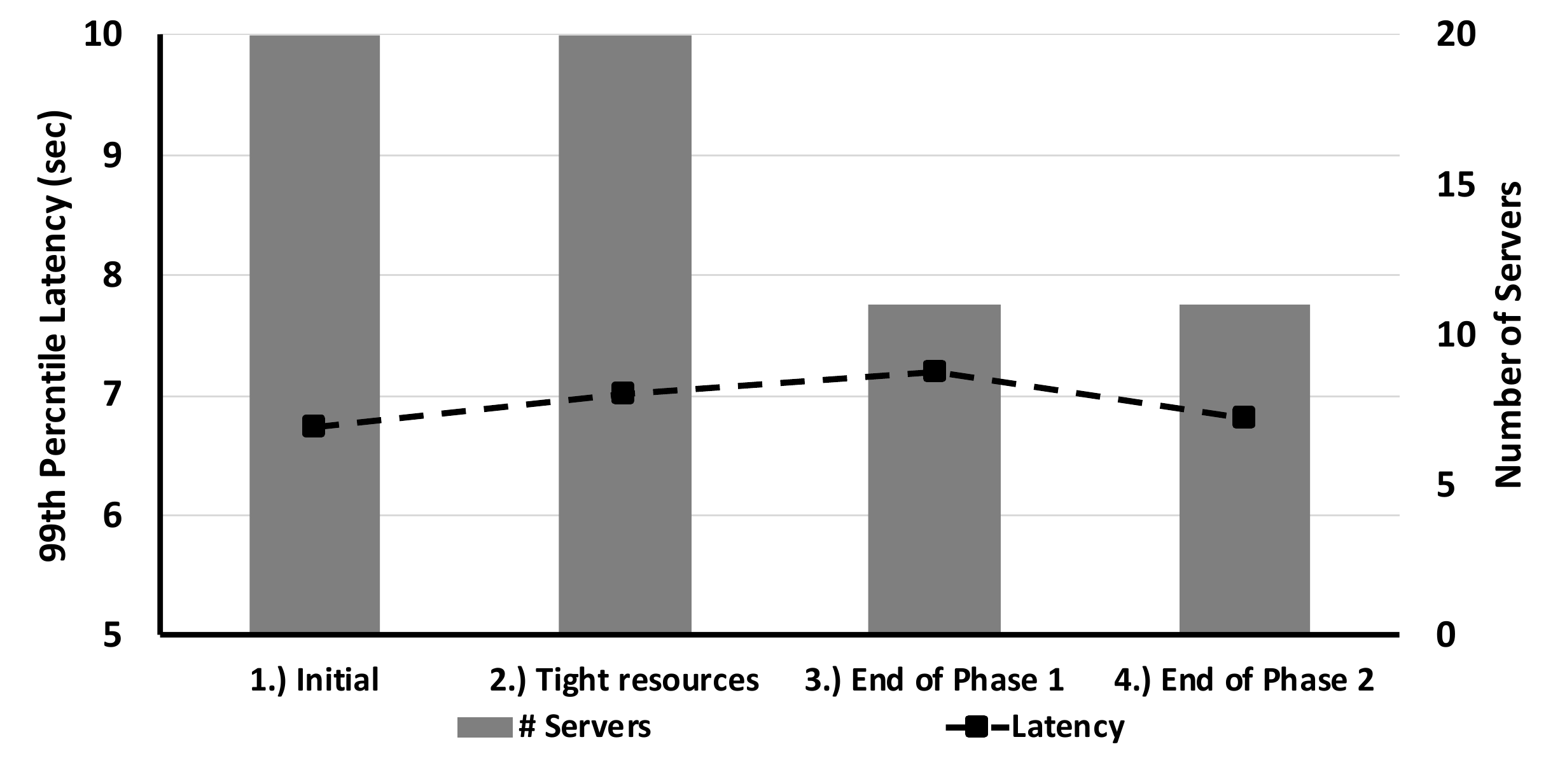}} 
\caption{Changes in  number of servers (bars) and 99 percentile response latency (lines) over the course of running \systemname end-to-end. We show both values at (1) the initial configuration; (2) after overprovisioned MRs have been reduced; (3) at the end of phase 1 which includes changes to placement; (4) after phase 2.}
\label{fig:e2e_perf}
\end{figure}

\subsection{End-to-end Experiments}
\label{subsection:e2e}
We start by applying \name to the applications in \S\ref{sec:application}. We show results for the MEAN stack in Figure~\ref{fig:mean-e2e}, the apartment application with mixed workload in Figure~\ref{fig:aptv2-e2e}, and the HotROD application in Figure \ref{fig:hotrod-e2e}. In these graphs, we show how performance (99th percentile latency) and server usage changes as \name runs. We show performance (line) and number of servers (bar) at four points: (1) the initial overprovisioned deployment, (2) after stressing and reducing each microservices resources (but before changing placement), (3) at the end of phase 1 (\S\ref{sec:clampdown}) after microservices have been packed onto fewer servers, and (4) at the end of the performance improvement phase (\S\ref{sec:climbing}). We use the performance of the initial deployment as our baseline, this is because \name's primary goal is to achieve the same performance with fewer resources. Additionally, finding an optimal deployment (\ie the global minima for latency) would require an exhaustive search which is infeasible for the applications we consider here.


In the three applications we evaluated, the initial deployment split resources evenly between the three microservice instances. After running \name, the MEAN stack achieved $33\%$ better performance with half the number of servers, the apartment application achieved $20\%$ better performance with $2.5\times$ fewer servers, and the HotROD application used half the servers with roughly the same performance level. In order to determine these final deployments, \systemname required roughly 13, 18, and 18 hours for MEAN stack, Apartment App, and HotRod App, respectively.\footnote{These times might seem long, but we run every trial up to 25 times in order to reduce fluctuations in the results.} While the exact improvements vary based on application, we see significant improvements in all for resource efficiency, and performance that is \textit{at least} as good as the initial deployment -- in a reasonable amount of time.

The graphs also show that the MEAN stack and the write heavy workload of the apartment application experience slight performance degradation in the initials stages of the process, but show overall improvements after phase 2. This is because the available network bandwidth between microservices increases when they are packed onto fewer machines, and this shifts the bottleneck (in this case to the CPU). The Performance Improvement Phase can hence correct for this problem through resource transfer. While \systemname was able to improve the HotROD application's resource efficiency, observe that performance does not improve in the Performance Improvement Phase. The HotROD application contains 3 separate microservices that are all compute bottlenecked; consequently, in the clampdown deployment, there was very little free resources for impacted MRs to grow into. Nevertheless, \systemname was able to maintain the performance of the application. This is in contrast to the Apartment Application where all three resources are bottlenecks for different microservices (CPU, Disk, and network); this diversity enabled performance improvements in the Performance Improvement Phase.
    
\subsection{\name Robustness}
\label{subsection:e2e_robustness}
Next we evaluate \name's robustness to variations in user input and deployment conditions.

\subsubsection{Robustness to Initial Deployment}
\label{subsubsection:robust_initialconfig}

Recall that \systemname identifies locally optimal resource allocations; thus the operator-provided initial deployment can have a significant impact on the outcome. In the end-to-end experiments shown above (\ref{subsection:e2e}), the initial deployments placed three microservice instances on each server, where each microservice instance has an equal split of the server's total resources. We look at three different sets of initial deployments. 1.) deployments where each microservice instance has been provisioned the maximal amount of resources for the instance type 2.) \emph{randomly generated} deployments and 3.) deployments that result from horizontal autoscaling. 

\textbf{Maximally Provisioned Initial Deployment:} In this initial deployment, each microservice instance was provisioned an entire EC2 VM instance. In doing so, \systemname continued to demonstrate effectiveness in reducing server usage. \systemname was able to reduce server usage by 75\% for the Apartment App and the HotRod App by 40\%. However, \systemname did not register performance improvements because IMRs were already maximally provisioned; during the performance improvement phase, there were no resources to transfer because the impacted MRs were already maximally provisioned. Compared to the initial deployments described in the end-to-end experiments (with 3 per machine), \systemname identifies a deployment that is more performant but also more expensive. For instance, for the HotRod application, starting from a maximally provisioned initial deployment results in a final deployment that requires 2.5x more servers, but also achieves 2.2x better performance. 

\textbf{Randomized Initial Deployment:}  Table \ref{tab:randomconfiguration} demonstrates four such random deployments with the Apartment App. Note that in all cases, \systemname was able to cut resource usage by 57\% -- all while either improving or maintaining performance. Interestingly, regardless of the initial performance, 3 out of 4 of the random initial deployments converged to the same local optima. In the one auspicious scenario where the initial random deployment resulted in improved performance, \systemname was able to converge back to that same performance. For brevity, we elide results from other applications.

\begin{table}[th]
\footnotesize
\centering
\resizebox{\columnwidth}{!}{
\begin{tabular}{@{}rrrrrr@{}}
\toprule
\multicolumn{1}{c}{\textbf{Random Dep. \#}} & \multicolumn{1}{c}{\textbf{\# Server Reduction}}  & \multicolumn{1}{c}{\textbf{Init. Perf. (s)}} & \multicolumn{1}{c}{\textbf{Fin. Perf. (s)}} & \multicolumn{1}{c}{\textbf {\% Improve}} \\ \midrule
1 & 57\% & 68.3 & 45.2 & 33.8\\
2 & 57\% & 38.2 & 38.7 & -0.01\\
3 & 57\% & 63.4 & 47.0 & 25.9\\
4 & 57\% & 58.5 & 47.9 & 18.1
\end{tabular}}
\caption{End-to-end runs of Apartment Application App with randomly generated initial deployments}
\label{tab:randomconfiguration}
\vspace{-0.2in}
\end{table}

\textbf{Horizontal Autoscaling: } In Section \ref{sec:additional}, we discussed how \systemname could be used jointly with dynamic workloads. In this section, we presented two possible scenarios. In this section, we step through the \systemname-based solution where increased workload results in performance degradation. 

Upon a workload change, resource allocations revert to being overprovisioned (\ie the initial deployment) and will horizontally scale until some threshold in the scaling policy is met. Suppose the operator decides that this (now stabilized) workload level occurs regularly; the operator can optionally offline deploy \systemname using that new workload level. Later, when that workload level is encountered again, \systemname can alter the resource allocations without any operator intervention. What is the performance and resource efficiency advantage of this approach over horizontal autoscaling? To this end, we deployed all three of our microservice applications on Kubernetes, and enabled horizontal autoscaling (HPA). HPA in Kubernetes implements a control loop that checks the average percentage utilization across all pods, and scales up if that threshold is exceeded. We ran our experiments across two separate utilization thresholds: 10\% and 90\%. 

To simulate a sudden burst in load, we increased the workload by a factor of 10x. Upon observing the performance degradation as a result of the increased workload, the horizontal autoscaler would first take over, switching to an overprovisioned resource allocation. Rather than falling back to this overprovisioned deployment everytime, the operator could record the increased workload and subsequently run \systemname on that higher workload. 

The comparison between the horizontal autoscaling approach and \systemname are in Table \ref{tab:horizontal_autoscaling}. Performance ranged from very slight decreases to 20\% increases, while the number of servers decreased between 12\% and 52\%. We omit the result of HotRod with 10\% scaling because it scaled aggressively beyond the capacity of our resources.

\begin{table}[t]
\centering
\footnotesize
\resizebox{\columnwidth}{!}{

\begin{tabular}{|l|c|c|c|c|c|}
\hline
\multicolumn{1}{|c|}{} & \multicolumn{1}{|c|}{} & \multicolumn{2}{c|}{w/o \systemname} & \multicolumn{2}{c|}{w/ \systemname} \\

\cline{3-6}
\multicolumn{1}{|c|}{Application} & \multicolumn{1}{|c|}{Scaling Thres} & Servers & Perf & Servers  & Perf\\
\hline
MEAN          & 90\% & 11   & 1668  & 6 & 1630 \\
Apartment App & 90\% & 21   & 7919  & 13 & 7615 \\
HotRod        & 90\% & 22   & 20781 & 18 & 21302 \\
MEAN          & 10\% & 29  & 1141 & 14 & 1188\\
Apartment App & 10\% & 25  & 7686 & 22 & 6158 \\
HotRod        & 10\% & --  & --   & --   & --\\
\hline
\end{tabular}}
\caption{Server Usage and Performance resulting from horizontal autoscaling, compared against use of \systemname. Horizontal Autoscaler scales when avg. utilization exceeds \textit{scaling thres}}
\label{tab:horizontal_autoscaling}
\vspace{-0.2in}
\end{table}

\subsubsection{Robustness to Noise}
\label{subsubsection:robustness_noise}


As we discussed in \S\ref{sec:additional}, a microservice's performance can be influenced by other colocated programs, including programs running in other VMs belonging to other tenants, impacting the measurements that \name relies on. We evaluate the impact of this by deploying our application containers alongside containers running computation at random intervals. This emulates the effect of a multitenant environment. The noisy neighbors run workloads that use all resources \ie CPU, disk, memory and network. To emulate multitenant environments, these neighboring  containers are pinned to a different core from the application under test.

Table \ref{tab:noise_table} shows the server reductions across deployments with and without noise. Server reductions remain unchanged while performance improvements varied somewhat (in both directions). Note that the results in this table are deployed on a different instance type (with more resources to facilitate the noisy neighbors) and should not be compared with earlier end-to-end experiments.

\begin{table}[t]
\footnotesize
\centering
\resizebox{\columnwidth}{!}{
\begin{tabular}{@{}lrrrr@{}}
\toprule
\multicolumn{1}{c}{\textbf{Application}} & \multicolumn{1}{c}{\textbf{\% Server Reduction}} & \multicolumn{1}{c} {\textbf{\% Performance Improvement}} \\ \midrule
Apartment   & 57      & 14          \\
Apartment (w/ noise) & 57    & 10 \\
HotRod          & 42      & 4.6     \\
HotRod (w/ noise) & 42   & 1   \\
MEAN       & 50      & 4    \\
MEAN (w/ noise) &  50  & 17 
\end{tabular}}
\caption{Comparing \systemname's efficacy towards server reduction and performance increase with and without noise.}
\label{tab:noise_table}
\vspace{-0.2in}
\end{table}

\begin{figure}[t]
     \centering
     \includegraphics[width=0.4\textwidth]{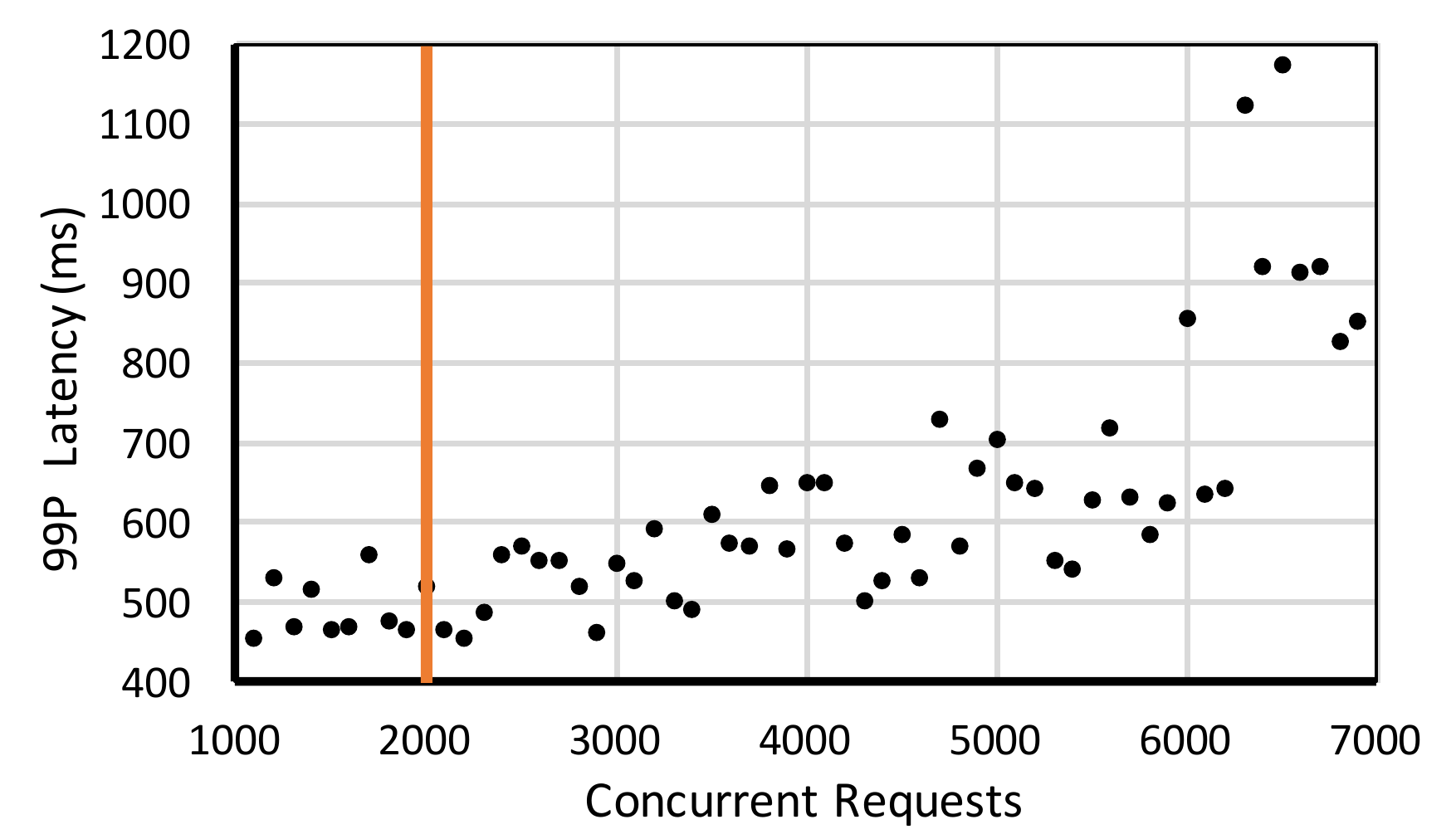}
     \caption{Robustness of MEAN App performance with additional concurrent requests. \systemname used a workload of 2000 requests, indicated by the dashed vertical line.}\label{fig:mean_robustness}
     \vspace{-.2in}
 \end{figure}

\subsubsection{Robustness to Workload Fluctuations}
\label{subsubsection:robustness_workload}
\name's mission is simple: given an initial deployment and a representative workload, \name tries to find a more efficient deployment (fewer servers) with at least as good performance. One might worry that after this process of reducing servers, the resulting deployment would be so highly tuned that it would become very sensitive to workload fluctuations. This would be bad, since fluctuations are inherent in customer-driven applications.
Fortunately, in our experience of running the applications we discuss here, the resulting performance is not highly sensitive to small workload fluctuations. We illustrate this with an example.

We took the MEAN application, and then altered the workload by increasing the rate of requests. Because the clients operate in closed request loops (request, then response, then new request), we merely increased the concurrency of this loop to issue several simultaneous requests. The original workload had a concurrency of 2000, and Figure \ref{fig:mean_robustness} shows how MEAN application performance changes as this level of concurrency is varied, while using the deployment computed by \name on the original workload. As shown in the graph, 99p latency grows gradually until the number of concurrent requests triples (from 2000 to 6000 concurrent requests), at which point the performance starts degrading far more rapidly. At this point, the operator would benefit from rerunning \systemname with a workload consisting of 6000 concurrent requests or more.

This fits with our general experience with \name that application performance does not suddenly degrade. One reason for this is that while the clampdown process produces a tight resource allocation, once these allocations are used to produce a new deployment, there are excess resources on each server because one typically cannot perfectly binpack the microservices into a minimal set of servers. In addition, applications are typically designed to degrade gracefully rather than fall off a cliff, and this might provide another reason for resilience.

\eat{In addition, recall that we assumed that the representative workload was supposed to be on the high-side, to directly provide some buffer from workload variations. Thus, between the inherently conservative nature of the representative workload, and the observed resilience against moderate load increases, we are less worried about \name being overly susceptible to workload fluctuations.}

\subsection{Resource Clampdown Phase Illustration}
\label{subsection:clampdown}

Now that we have examined \name's overall impact in terms of server reduction and application performance, we focus in on the two phases of \systemname to see where these improvements come from. In this subsection, we demonstrate the resource efficiency gains and runtime of just Resource Clampdown Phase. Table \ref{tab:clampdown_table} details the server savings as well as the Clampdown runtime. In keeping with the previous sections, we provide two initial deployment configurations: one container per machine, and three containers per machine. Only one initial deployment is shown for MEAN stack, since the MEAN stack only has four microservice instances. We find that in all cases resource utilization reduces at the end of this phase, and that across applications this phase took between 30 minutes and 7 hours. This time depends on several factors including the input deployment and the chosen workload, and can hence be reduced by having the operator make appropriate choices.

\begin{table}[t]
\footnotesize
\centering
\resizebox{\columnwidth}{!}{
\begin{tabular}{@{}lrrrr@{}}
\toprule
\multicolumn{1}{c}{\textbf{Application}} & \multicolumn{1}{c}{\textbf{Initial Servers}} & \multicolumn{1}{c}{\textbf{Final Servers}} & \multicolumn{1}{c}{\textbf{Iterations}}  & \multicolumn{1}{c}{\textbf{Clampdown Hours}}\\ \midrule
Apartment App Write  & 21       & 3     & 3  & 1     \\
Apartment App Write    & 7    &  3       &  4 & 1 \\
Apartment App Mix       & 21       & 5     & 8  & 3.5     \\
Apartment App Mix      & 7       & 4     & 4   & 1.75 \\
HotRod                 & 21      & 11     & 9   & 5 \\ 
HotRod                 & 7       & 4     & 6   &  4.4 \\
MEAN Stack             & 4       & 2     & 4   & 1
\end{tabular}}
\caption{Running just Resource Clampdown on applications across two different initial deployment settings.}
\label{tab:clampdown_table}
\vspace{-0.2in}
\end{table}

\subsection{Performance Improvement Phase Illustration}
\label{subsection:perf_improvement}

In this subsection, we evaluate the second phase of \systemname: Performance Improvement. Like the previous section, we start by providing an overview of its performance gains and runtime. Next, we go a step deeper into the two primary mechanisms that underlie the Performance Improvement Phase. First, we evaluate the effectiveness of resource stressing for IMR identification. We then provide an example of how resource transfer works.

\begin{table}[t]
\footnotesize
\centering
\resizebox{\columnwidth}{!}{
\begin{tabular}{@{}lrrrr@{}}
\toprule
\multicolumn{1}{c}{\textbf{Application}} & \multicolumn{1}{c}{\textbf{\% Improve}} & \multicolumn{1}{c}{\textbf{Time (hr)}}\\ \midrule
Apartment App Mix,99p latency       & 26       & 11   \\
Apartment App Mix, 50p latency      & 30.4     & 16   \\
MEAN Stack, 99p latency             & 65.9     & 9   \\
MEAN Stack, 50p latency             & 45       & 10   \\
HotRod, 99p latency                 & 6        & 12   \\
\end{tabular}}
\caption{Overall Performance gains from gradient step in performance hill climbing.}
\label{tab:hillclimb_table1}
\vspace{-0.2in}
\end{table}

\subsubsection{Performance Improvement Phase: results}
\label{subsubsection:performance_improvement_results}
We start by providing overall results from applying Performance Hill Climbing to these applications, before evaluating the individual steps in this phase. Table~\ref{tab:hillclimb_table1} shows the performance improvements in this phase, showing that Performance Hill Climbing in these cases improves application performance by between 17\% to 66\%. While achieving this improvement takes several hours for some applications, this is determined by the performance metric and workload selected by the operator, and hence depends on the setting in which \systemname is used.

\begin{table}[th]
\footnotesize
\centering
\resizebox{\columnwidth}{!}{
\begin{tabular}{@{}lrrrr@{}}
\toprule
\multicolumn{1}{c}{\textbf{Application}} & \multicolumn{1}{c}{\textbf{Single MR}} &  \multicolumn{1}{c}{\textbf{\% Improve}}  & \multicolumn{1}{c}{\textbf{Most impacted?}}\\ \midrule
Apartment App Write  & Node, CPU            & 30.4  &Yes     \\
Apartment App Mixed  & Node, CPU           & 16.1  &Yes     \\
HotROD App.          & Api, CPU         & 16.1          & Yes \\
MEAN Stack           & Node, CPU          & 55.5   &Yes 
\end{tabular}}
\caption{Effect of Provisioning more resources to the Most Impacted MR on the first iteration of the Performance Hill Climbing Phase.}

\label{tab:mimr}
\vspace{-0.2in}
\end{table}

\begin{figure*}[th]
\centering
\minipage{0.45\textwidth}
    \includegraphics[width=\textwidth]{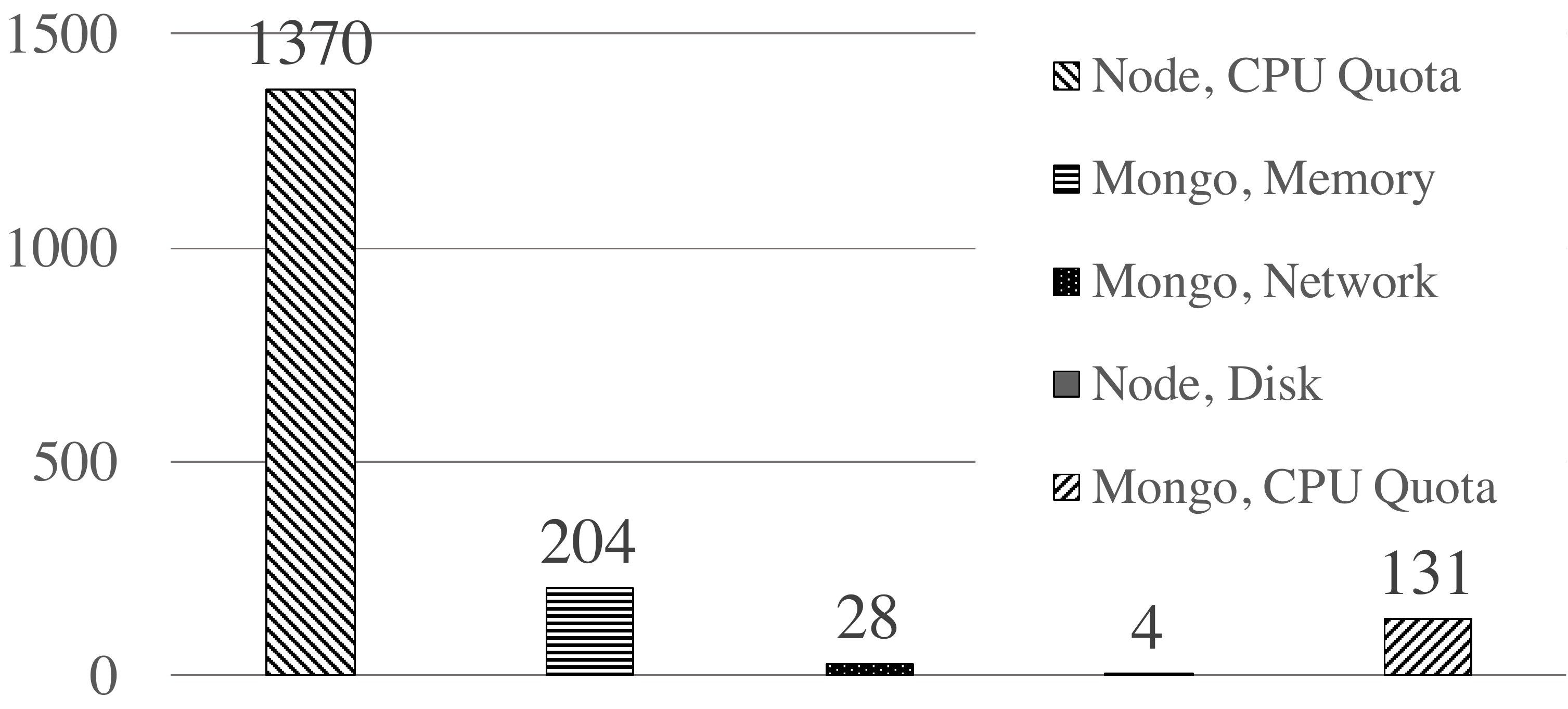}
    \caption{MEAN Stack Performance Degradation upon Single Resource Gradient ($99^{th}$ percentile latency)}\label{fig:mean-mr-stress}
    \vspace{-0.1in}
\endminipage\hfill
\minipage{0.45\textwidth}
    \includegraphics[width=\textwidth]{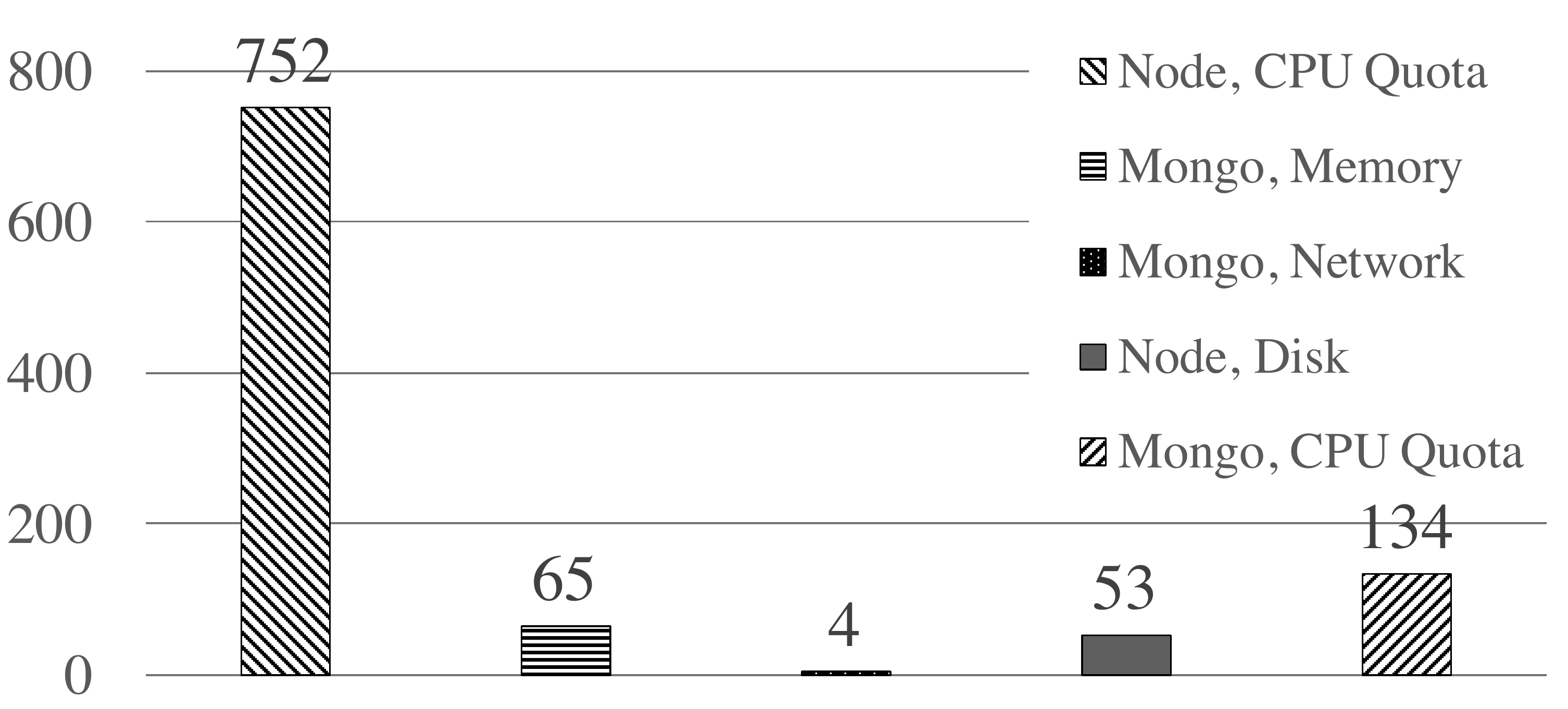}
    \caption{MEAN Stack Performance Improvement upon actual improvement resources ($99^{th}$ percentile latency)}\label{fig:mean-mr-actual}
    \vspace{-0.1in}
\endminipage\hfill
\end{figure*}

\subsubsection{Performance Improvement Phase: Stressing Effectiveness}
\label{subsection:mechanism}

We dive a level deeper and explore the inner workings of \systemname, starting with the fundamental mechanism of \systemname: resource stressing. The ability to determine the relative sensitivity of an MR underlies \systemname's effectiveness. 
Since \systemname uses stressing to infer something about each MR, false positives are possible in the Performance Improvement Phase. While we empirically have found that \systemname does not require an exact, complete ordering of MR sensitivity, the algorithm revolves around two basic tasks that are important to get right. The first is identifying the most impacted MR, as it is the most direct way of improving \systemname performance in the Performance Improvement Phase. The second is to  distinguish impacted MRs from non-impacted MRs, since we only want to transfer resources to impacted MRs. In this subsection, we illustrate how effectively \systemname accomplishes these two goals for the applications we tested.

First, we demonstrate that resource stressing is an effective way of identifying the most impacted MR. Table \ref{tab:mimr} shows that in every application that we tested, \systemname was able to identify the most impacted MR correctly. We confirmed that the MR was indeed the most impacted by exhaustively increasing each MR's resource allocation and measuring the resulting performance.\footnote{Note that we can do this kind of resource increase for this special test, but when running \name such increases are not possible because they typically violate the resource constraints on the server.}

Next, we look at \systemname's ability to positively identify an impacted MR. We illustrate IMR identification with an example using the MEAN stack. Figure \ref{fig:mean-mr-stress} shows the measured end-to-end performance resulting from stressing the a single MR. Note that we omit some MRs from the figure for clarity; the MRs we removed did not impact performance when resource allocation was decreased. Based on this result of this gradient, \systemname would identifies following MRs as being IMRs: (Node, cpu), (Mongo, memory), and (Mongo, cpu). When compared to the the actual performance improvements for the same MRs (Figure~\ref{fig:mean-mr-actual}), we see that provisioning more resources to those exact three MRs results in performance gains. The other MRs -- correctly identified as NIMRs -- did not result in significant performance improvements when the resource allocation was increased. For the MEAN stack, the stressing mechanism positively identified all the IMRs. Additionally, note that provisioning more resources to the most impacted MR causes end-to-end performance to improve by 752 ms (which is nearly 30\%). It is important to re-emphasize here that the Performance Improvement Phase makes no guarantees about the exact \textit{magnitude} of the performance improvement upon the mitigation of the bottleneck. 

We observed similar behavior among the Hotrod and Apartment Apps.

\subsubsection{Performance Improvement Phase: Resource Transfer Example}
\label{subsubsection:resource_stealing}

Now that we have shown that \systemname can positively identify IMRs, we discuss concretely how resource transfer works.
The gradient phase improves performance by transferring resources between less impacted MRs and more impacted MRs. Transferring resources in this way should work well if everything was cleanly linear in behavior. However, this assumption does not hold in practice, and simultaneously taking two actions (\eg moving resources from an NIMR to an IMR) can yield unexpected behavior.

To evaluate how well this works in practice, we observed the ability of \systemname to transfer resources in various setups. Within the Apartment App, we zoomed into the ELK stack, which consists of three microservices: Elasticsearch, Logstash, and Kibana. CPU-Quota was initially evenly distributed between Elasticsearch and Kibana. The baseline performance under this configuration was measured to be 31.49 seconds. In the first iteration, \systemname identifies Elasticsearch CPU-Quota as the most impacted MR. Upon reducing the resource allocation of $(Kibana, cpu-quota)$ by 25\% and increasing $(Elasticsearch, cpu-quota)$ by the same amount, performance improved to 28.48 seconds, equivalent to the performance exhibited simply by improving $(Elasticsearch, cpu-quota)$ by 25\% whilst leaving $(Kibana, cpu-quota)$ to be constant. We observed similar behavior for the MEAN stack.

While this does not cover the space of all application behaviors (even within the applications that we explored), thus far we have not seen an application where transfering resources from a correctly identified non-impacted MR and provisioning it to a correctly identified most-impacted MR has resulted in undesirable behavior. In the event of this undesirable behavior, \systemname is capable of detecting it. Once detected, the \textit{backtrack step} would simply take over and ultimately undo the resource transfer.

\subsection{Additional Issue: Effect of Interference}
In this section, we dive into the issue of interference, which we discussed in Section \ref{sec:additional} (Additional Issues). In particular, we illustrate how \systemname copes with interference on the HotRod application, the only application we tested where placement contributed to the application performance. This is because the HotRod applications exhibits a significant number of compute-impacted MRs. \systemname empirically detects interference by measuring the clamped down performance with the original placement and compares it with the performance on the new compacted placement. For the HotRod application, performance degraded by 12\% -- despite having the same resource allocations. 

Recall that \systemname reduces the likelihood of new interference by 1.) preserving colocated microservice instances from the initial deployment and 2.) exclusively assigning core(s) to those microservices. Upon applying affinity-based placements and core pinning, performance degradation upon clampdown decreased to 2\%. With this mechanism, more compute resources were utilized, but the number of packed nodes remained constant. Ultimately, for this version of the HotRod application, this simple mechanism improves performance after the clampdown phase at no additional cost to the operator.  There are clear limitations to this approach, especially when all compute MRs are \textit{similarly} impacted. Our solution here does not attempt to exhaustively eliminate container interference; other projects have addressed this problem extensively~\cite{Lo:2015:HIR:2749469.2749475,Tootoonchian:2018:RES:3307441.3307466}. Nevertheless, in the only case where our applications encountered interference, the combination of core pinning and invariant placement sufficiently mitigated the effect of placement on performance.

\end{outline}

\eat{\textbf{Software Configurations as \systemname Inputs}: Some services, \eg Spark, require that configuration be updated (\eg by increasing heap size) to take advantage of additional resources. The complexity of these configuration changes varies across services: some services such as Nginx allow configuration to be updated at runtime, while others such as Spark need to be restarted before they can make use of additional resources. The user is required to provide this configuration as input. This requirement to change configuration adds noticeable overheads when getting started with and deploying \systemname. As a result \systemname is better suited to handling services that are designed to dynamically adapt to using additional resources at runtime. In our own experiences running \systemname, most reconfiguration programs are expressible as a simple Bash script. We however note that several services (e.g., Postgres \cite{postgresql}, NGINX) already provide tools (\eg \texttt{pgtune}) and APIs or dynamic configuration. Furthermore, tools such as Kubernetes Helm \cite{helm} have begun developing tools that apply across services. We plan to more fully investigate the challenges of integrating with automatic configuration tools in future work.\\}

\section{Related Work}
Recent work to determine resource allocations has largely focused on provisioning resources for \textit{individual microservices}, rather than considering complex dependencies in end-to-end heterogeneous microservice applications. These projects can take either a data-driven or domain specific approach. Examples of work that adopt the data-driven approach include Paragon~\cite{Delimitrou2013ParagonQS} and Quasar~\cite{Delimitrou:2014:QRQ:2541940.2541941} that rely on microarchitectural details to reduce colocation overheds. While these works do not make any assumptions about the application, and can account for behaviours such as noisy neighbors (which are ignored by \systemname), they cannot scale to applications containing several microservices. Domain specific approaches include work such as Ernest~\cite{Venkataraman:2016:EEP:2930611.2930635}, CherryPick~\cite{Alipourfard2017CherryPickAU} and Paris~\cite{Yadwadkar:2017} which make strong assumptions about application structure (\eg assuming map-reduce like data parallel frameworks) or that the application is comprised of homogeneous services.
Other approaches to the resource allocation problem have relied on the use of instrumentation and custom tooling to infer application dependencies. This approach adds performance overheads and increases system complexity complexity. Sieve~\cite{DBLP:journals/corr/abs-1709-06686} considers performance across multiple microservices, but requires instrumentation like \textit{sysdig} or \textit{dtrace}, with overhead of up to 100\% in the worst case~\cite{dtrace}. Additionally, these tools suffer from causation ambiguity and leads to high rates of false positives~\cite{Lee2013HighAA, Ji2017RAINRA}. Other proposed systems allow operators to infer performance behaviors of various systems, but they largely require significant modifications to the hypervisor or to the application under test \cite{Gupta:2011:DTD:1963559.1963560, Gupta:2005:IBT:1095810.1118605} or only explore performance improvements along a single resource dimension (e.g., network) \cite{Gupta:2005:IBT:1095810.1118605,Pan2005SHRiNKAM,Bucy2008TheDS}.
Other works have suggested improving application performance through profiling programs and optimizing code \cite{Curtsinger:2015:COF:2815400.2815409, Miller1995ThePP, Burtscher2010PerfExpertAE}; these provide a different set of knobs from impacted MR allocations, and thus can be jointly used with \systemname to improve application performance. Past proposals on resource scheduling \cite{Ghodsi:2011:DRF:1972457.1972490, Mao:2016:RMD:3005745.3005750, Grandl2014MultiresourcePF} assume that the administrator provides resource requirements as input; thus we view these work as complimentary.

\section{Conclusion}
It is important to consider \name relative to its goals. \name was not intended to provide optimal performance, nor optimal efficiency, nor provable guarantees, nor to work well with arbitrarily varying workloads. Instead, it was designed to be easy-to-use (operators need only provide an initial deployment, a representative workload, and a performance metric) and generally applicable (it does not make any assumptions about the nature of the application) tool that reduces the number of servers needed to deploy applications while maintaining (or improving) performance. We believe \name achieves this goal, and we are not aware of any other tool that does so. However, only widespread use of \name will allow us to fully understand its limitations, which hopefully will lead to further improvements.

\section{Acknowledgments}
We thank members of the NetSys Lab at UC Berkeley, and Shivaram Venkatraman for their comments on this work. This work was funded in part by funding from VMware, Intel and NSF grant 1817116.
\bibliographystyle{plainnat}
\bibliography{tbot}

\begin{thebibliography}{48}
\providecommand{\natexlab}[1]{#1}
\providecommand{\url}[1]{\texttt{#1}}
\expandafter\ifx\csname urlstyle\endcsname\relax
  \providecommand{\doi}[1]{doi: #1}\else
  \providecommand{\doi}{doi: \begingroup \urlstyle{rm}\Url}\fi

\bibitem[gor()]{goreplay}
Goreplay.
\newblock \url{https://goreplay.org/}, retrieved 1/13/2020.

\bibitem[hel()]{helm}
Helm -- the kubernetes package manager.
\newblock \url{https://helm.sh/}, retrieved 9/17/2018.

\bibitem[tod()]{todo-app}
Node todo-app.
\newblock \url{https://github.com/kelda/node-todo}, retrieved 1/15/2020.

\bibitem[ab()]{apachebench}
ab.
\newblock {ab - Apache HTTP server benchmarking tool}.
\newblock \url{https://httpd.apache.org/docs/2.4/programs/ab.html}, retrieved
  10/27/2017.

\bibitem[Alipourfard et~al.(2017)Alipourfard, Liu, Chen, Venkataraman, Yu, and
  Zhang]{Alipourfard2017CherryPickAU}
Omid Alipourfard, Hongqiang~Harry Liu, Jianshu Chen, Shivaram Venkataraman,
  Minlan Yu, and Ming Zhang.
\newblock Cherrypick: Adaptively unearthing the best cloud configurations for
  big data analytics.
\newblock In \emph{NSDI}, 2017.

\bibitem[{Amazon Web Services}()]{aws:elb}
{Amazon Web Services}.
\newblock {Elastic Load Balancing and Amazon EC2 Auto Scaling}.
\newblock
  \url{https://docs.aws.amazon.com/autoscaling/ec2/userguide/autoscaling-load-balancer.html}.

\bibitem[Bucy et~al.(2008)Bucy, Schindler, Schlosser, and
  Ganger]{Bucy2008TheDS}
John~S. Bucy, Jiri Schindler, Steven~W. Schlosser, and Gregory~R. Ganger.
\newblock \emph{The DiskSim Simulation Environment Version 4.0 Reference
  Manual}.
\newblock CMU PDL, 2008.

\bibitem[Burns et~al.(2016)Burns, Grant, Oppenheimer, Brewer, and
  Wilkes]{Burns2016BorgOA}
Brendan Burns, Brian Grant, David Oppenheimer, Eric Brewer, and John Wilkes.
\newblock Borg, omega, and kubernetes.
\newblock \emph{Commun. ACM}, 59:\penalty0 5, 2016.

\bibitem[Burtscher et~al.(2010)Burtscher, Kim, Diamond, McCalpin, Koesterke,
  and Browne]{Burtscher2010PerfExpertAE}
Martin Burtscher, Byoung-Do Kim, Jeffrey~R. Diamond, John~D. McCalpin, Lars
  Koesterke, and James~C. Browne.
\newblock Perfexpert: An easy-to-use performance diagnosis tool for hpc
  applications.
\newblock In \emph{SC}, 2010.

\bibitem[Curtsinger and Berger(2015)]{Curtsinger:2015:COF:2815400.2815409}
Charlie Curtsinger and Emery~D. Berger.
\newblock Coz: Finding code that counts with causal profiling.
\newblock In \emph{SOSP}, 2015.

\bibitem[datadog()]{datadog}
datadog.
\newblock Datadog: Cloud monitoring as a service.
\newblock \url{https://www.datadoghq.com/}.

\bibitem[Dean and Barroso(2013)]{Dean2013TheTA}
Jeffrey Dean and Luiz~Andr{\'e} Barroso.
\newblock The tail at scale.
\newblock \emph{Commun. ACM}, 56:\penalty0 74--80, 2013.

\bibitem[Delimitrou and Kozyrakis(2013)]{Delimitrou2013ParagonQS}
Christina Delimitrou and Christoforos~E. Kozyrakis.
\newblock Paragon: Qos-aware scheduling for heterogeneous datacenters.
\newblock In \emph{ASPLOS}, 2013.

\bibitem[Delimitrou and Kozyrakis(2014)]{Delimitrou:2014:QRQ:2541940.2541941}
Christina Delimitrou and Christos Kozyrakis.
\newblock Quasar: Resource-efficient and qos-aware cluster management.
\newblock In \emph{Proceedings of the 19th International Conference on
  Architectural Support for Programming Languages and Operating Systems},
  ASPLOS '14, pages 127--144, New York, NY, USA, 2014. ACM.
\newblock ISBN 978-1-4503-2305-5.
\newblock \doi{10.1145/2541940.2541941}.
\newblock URL \url{http://doi.acm.org/10.1145/2541940.2541941}.

\bibitem[Ghodsi et~al.(2011)Ghodsi, Zaharia, Hindman, Konwinski, Shenker, and
  Stoica]{Ghodsi:2011:DRF:1972457.1972490}
Ali Ghodsi, Matei Zaharia, Benjamin Hindman, Andy Konwinski, Scott Shenker, and
  Ion Stoica.
\newblock Dominant resource fairness: Fair allocation of multiple resource
  types.
\newblock In \emph{NSDIR}, 2011.

\bibitem[{Google Cloud}()]{google:autoscale}
{Google Cloud}.
\newblock {Google Cloud: Autoscaling groups of instances}.
\newblock \url{https://cloud.google.com/compute/docs/autoscaler}.

\bibitem[Grandl et~al.(2014)Grandl, Ananthanarayanan, Kandula, Rao, and
  Akella]{Grandl2014MultiresourcePF}
Robert Grandl, Ganesh Ananthanarayanan, Srikanth Kandula, Sriram Rao, and
  Aditya Akella.
\newblock Multi-resource packing for cluster schedulers.
\newblock In \emph{SIGCOMM}, 2014.

\bibitem[Gregg(2011)]{dtrace}
Brendan Gregg.
\newblock {dtrace pid provider overhead}.
\newblock
  \url{http://dtrace.org/blogs/brendan/2011/02/18/dtrace-pid-provider-overhead/}
  retrieved 05/01/2018, 2011.

\bibitem[Gupta()]{microservicedesignpattern}
Arun Gupta.
\newblock Microservice design patterns.
\newblock \url{http://blog.arungupta.me/microservice-design-patterns/},
  retrieved 9/17/2018.

\bibitem[Gupta et~al.(2005)Gupta, Yocum, McNett, Snoeren, Vahdat, and
  Voelker]{Gupta:2005:IBT:1095810.1118605}
Diwaker Gupta, Kenneth Yocum, Marvin McNett, Alex~C. Snoeren, Amin Vahdat, and
  Geoffrey~M. Voelker.
\newblock To infinity and beyond: Time warped network emulation.
\newblock In \emph{SOSP}, 2005.

\bibitem[Gupta et~al.(2011)Gupta, Vishwanath, McNett, Vahdat, Yocum, Snoeren,
  and Voelker]{Gupta:2011:DTD:1963559.1963560}
Diwaker Gupta, Kashi~Venkatesh Vishwanath, Marvin McNett, Amin Vahdat, Ken
  Yocum, Alex Snoeren, and Geoffrey~M. Voelker.
\newblock Diecast: Testing distributed systems with an accurate scale model.
\newblock \emph{ACM Transactions on Computer Systems}, 29\penalty0
  (2):\penalty0 4:1--4:48, May 2011.
\newblock ISSN 0734-2071.
\newblock \doi{10.1145/1963559.1963560}.
\newblock URL \url{http://doi.acm.org/10.1145/1963559.1963560}.

\bibitem[Hagberg et~al.(2008)Hagberg, Schult, and Swart]{networkx}
Aric~A. Hagberg, Daniel~A. Schult, and Pieter~J. Swart.
\newblock Exploring network structure , dynamics , and function using networkx.
\newblock In \emph{SciPy}, 2008.

\bibitem[Hubert(2001)]{tc_man}
Bert Hubert.
\newblock {tc(8)}.
\newblock Linux man page -- iproute2, 2001.

\bibitem[Jackson()]{kelda}
Ethan Jackson.
\newblock Kelda: An approachable way to deploy the cloud.
\newblock \url{https://github.com/kelda/kelda}, retrieved 9/17/2018.

\bibitem[Jalaparti et~al.(2013)Jalaparti, Bodik, Kandula, Menache, Rybalkin,
  and Yan]{jalaparti2013speeding}
Virajith Jalaparti, Peter Bodik, Srikanth Kandula, Ishai Menache, Mikhail
  Rybalkin, and Chenyu Yan.
\newblock Speeding up distributed request-response workflows.
\newblock \emph{SIGCOMM}, 2013.

\bibitem[Ji et~al.(2017)Ji, Lee, Downing, Wang, Fazzini, Kim, Orso, and
  Lee]{Ji2017RAINRA}
Yang Ji, Sangho Lee, Evan Downing, Weiren Wang, Mattia Fazzini, Taesoo Kim,
  Alessandro Orso, and Wenke Lee.
\newblock Rain: Refinable attack investigation with on-demand inter-process
  information flow tracking.
\newblock In \emph{CCS}, 2017.

\bibitem[Jyothi et~al.(2016)Jyothi, Curino, Menache, Narayanamurthy, Tumanov,
  Yaniv, Mavlyutov, Goiri, Krishnan, Kulkarni, and Rao]{Jyothi2016MorpheusTA}
Sangeetha~Abdu Jyothi, Carlo Curino, Ishai Menache, Shravan~M. Narayanamurthy,
  Alexey Tumanov, Jonathan Yaniv, Ruslan Mavlyutov, I{\~n}igo Goiri, Subru
  Krishnan, Janardhan Kulkarni, and Sriram Rao.
\newblock Morpheus: Towards automated slos for enterprise clusters.
\newblock In \emph{OSDI}, 2016.

\bibitem[Kumar et~al.(2016)Kumar, Ananthanarayanan, Ratnasamy, and
  Stoica]{cedar}
Gautam Kumar, Ganesh Ananthanarayanan, Sylvia Ratnasamy, and Ion Stoica.
\newblock Hold 'em or fold 'em? aggregation queries under performance
  variations.
\newblock In \emph{EuroSys}, 2016.

\bibitem[Lee et~al.(2013)Lee, Zhang, and Xu]{Lee2013HighAA}
Kyu~Hyung Lee, Xiangyu Zhang, and Dongyan Xu.
\newblock High accuracy attack provenance via binary-based execution partition.
\newblock In \emph{NDSS}, 2013.

\bibitem[Lo et~al.(2015)Lo, Cheng, Govindaraju, Ranganathan, and
  Kozyrakis]{Lo:2015:HIR:2749469.2749475}
David Lo, Liqun Cheng, Rama Govindaraju, Parthasarathy Ranganathan, and
  Christos Kozyrakis.
\newblock Heracles: Improving resource efficiency at scale.
\newblock In \emph{Proceedings of the 42Nd Annual International Symposium on
  Computer Architecture}, ISCA '15, pages 450--462, New York, NY, USA, 2015.
  ACM.
\newblock ISBN 978-1-4503-3402-0.
\newblock \doi{10.1145/2749469.2749475}.
\newblock URL \url{http://doi.acm.org/10.1145/2749469.2749475}.

\bibitem[Mao et~al.(2016)Mao, Alizadeh, Menache, and
  Kandula]{Mao:2016:RMD:3005745.3005750}
Hongzi Mao, Mohammad Alizadeh, Ishai Menache, and Srikanth Kandula.
\newblock Resource management with deep reinforcement learning.
\newblock In \emph{HotNets}, 2016.

\bibitem[Marathe et~al.(2017)Marathe, Anirudh, Jain, Bhatele, Thiagarajan,
  Kailkhura, Yeom, Rountree, and Gamblin]{10.1145/3126908.3126969}
Aniruddha Marathe, Rushil Anirudh, Nikhil Jain, Abhinav Bhatele, Jayaraman
  Thiagarajan, Bhavya Kailkhura, Jae-Seung Yeom, Barry Rountree, and Todd
  Gamblin.
\newblock Performance modeling under resource constraints using deep transfer
  learning.
\newblock In \emph{Proceedings of the International Conference for High
  Performance Computing, Networking, Storage and Analysis}, SC ’17, New York,
  NY, USA, 2017. Association for Computing Machinery.
\newblock ISBN 9781450351140.
\newblock \doi{10.1145/3126908.3126969}.
\newblock URL \url{https://doi.org/10.1145/3126908.3126969}.

\bibitem[{Microsoft Azure}()]{azure:autoscale}
{Microsoft Azure}.
\newblock {Azure Autoscale}.
\newblock \url{https://azure.microsoft.com/en-us/features/autoscale/}.

\bibitem[Miller et~al.(1995)Miller, Callaghan, Cargille, Hollingsworth, Irvin,
  Karavanic, Kunchithapadam, and Newhall]{Miller1995ThePP}
Barton~P. Miller, Mark~D. Callaghan, Jonathan~M. Cargille, Jeffrey~K.
  Hollingsworth, R.~Bruce Irvin, Karen~L. Karavanic, Krishna Kunchithapadam,
  and Tia Newhall.
\newblock The paradyn parallel performance measurement tool.
\newblock \emph{IEEE Computer}, 28:\penalty0 37--46, 1995.

\bibitem[Pan et~al.(2005)Pan, Prabhakar, Psounis, and Wischik]{Pan2005SHRiNKAM}
Rong Pan, Balaji Prabhakar, Konstantinos Psounis, and Damon Wischik.
\newblock Shrink: a method for enabling scaleable performance prediction and
  efficient network simulation.
\newblock \emph{IEEE/ACM Transactions on Networking}, 13:\penalty0 975--988,
  2005.

\bibitem[redhat()]{redhat}
redhat.
\newblock Placing pods relative to other pods using affinity and anti-affinity
  rules.
\newblock
  \url{https://docs.openshift.com/container-platform/4.1/nodes/scheduling/nodes-scheduler-pod-affinity.html},
  retrieved 9/16/2019.

\bibitem[redhatcgroups()]{redhat:cgroups}
redhatcgroups.
\newblock {Redhat Resource Management Guide}.
\newblock https://goo.gl/hiFDD7.

\bibitem[Rensin(2015)]{kubernetes}
David~K. Rensin.
\newblock \emph{Kubernetes - Scheduling the Future at Cloud Scale}.
\newblock O'Reilly Media, 2015.
\newblock URL \url{http://www.oreilly.com/webops-perf/free/kubernetes.csp}.

\bibitem[Richardson()]{microservicedesignpattern2}
Chris Richardson.
\newblock Pattern: Microservice architecture.
\newblock \url{https://microservices.io/patterns/microservices.html}, retrieved
  9/17/2018.

\bibitem[Shkuro()]{hotrodapp}
Yuri Shkuro.
\newblock Take opentracing for a hotrod ride.
\newblock
  \url{https://medium.com/opentracing/take-opentracing-for-a-hotrod-ride-f6e3141f7941},
  retrieved 9/17/2018.

\bibitem[{Suresh} et~al.(2019){Suresh}, {Loff}, {Kalim}, {Narodytska},
  {Ryzhyk}, {Gamage}, {Oki}, {Lokhandwala}, {Hira}, and
  {Sagiv}]{2019arXiv190903130S}
Lalith {Suresh}, Joao {Loff}, Faria {Kalim}, Nina {Narodytska}, Leonid
  {Ryzhyk}, Sahan {Gamage}, Brian {Oki}, Zeeshan {Lokhandwala}, Mukesh {Hira},
  and Mooly {Sagiv}.
\newblock {Automating Cluster Management with Weave}.
\newblock \emph{arXiv e-prints}, art. arXiv:1909.03130, Sep 2019.

\bibitem[Thalheim et~al.(2017)Thalheim, Rodrigues, Akkus, Bhatotia, Chen,
  Viswanath, Jiao, and Fetzer]{DBLP:journals/corr/abs-1709-06686}
J{\"{o}}rg Thalheim, Antonio Rodrigues, Istemi~Ekin Akkus, Pramod Bhatotia,
  Ruichuan Chen, Bimal Viswanath, Lei Jiao, and Christof Fetzer.
\newblock Sieve: Actionable insights from monitored metrics in microservices.
\newblock \emph{CoRR}, abs/1709.06686, 2017.
\newblock URL \url{http://arxiv.org/abs/1709.06686}.

\bibitem[{Todd Hoff}()]{uber}
{Todd Hoff}.
\newblock {Lessons Learned From Scaling Uber To 2000 Engineers, 1000 Services,
  And 8000 Git Repositories}.
\newblock \url{https://goo.gl/1MRvoT}, retrieved 01/21/2017.

\bibitem[{Tony Mauro}()]{netflix}
{Tony Mauro}.
\newblock {Adopting Microservices at Netflix: Lessons for Architectural
  Design}.
\newblock \url{https://goo.gl/DyrtvI}, retrieved 01/21/2017.

\bibitem[Tootoonchian et~al.(2018)Tootoonchian, Panda, Lan, Walls, Argyraki,
  Ratnasamy, and Shenker]{Tootoonchian:2018:RES:3307441.3307466}
Amin Tootoonchian, Aurojit Panda, Chang Lan, Melvin Walls, Katerina Argyraki,
  Sylvia Ratnasamy, and Scott Shenker.
\newblock Resq: Enabling slos in network function virtualization.
\newblock In \emph{Proceedings of the 15th USENIX Conference on Networked
  Systems Design and Implementation}, NSDI'18, pages 283--297, Berkeley, CA,
  USA, 2018. USENIX Association.
\newblock ISBN 978-1-931971-43-0.
\newblock URL \url{http://dl.acm.org/citation.cfm?id=3307441.3307466}.

\bibitem[Veeraraghavan et~al.(2016)Veeraraghavan, Meza, Chou, Kim, Margulis,
  Michelson, Nishtala, Obenshain, Perelman, and
  Song]{Veeraraghavan2016KrakenLL}
Kaushik Veeraraghavan, Justin Meza, David Chou, Wonho Kim, Sonia Margulis,
  Scott Michelson, Rajesh Nishtala, Daniel Obenshain, Dmitri Perelman, and
  Yee~Jiun Song.
\newblock Kraken: Leveraging live traffic tests to identify and resolve
  resource utilization bottlenecks in large scale web services.
\newblock In \emph{OSDI}, 2016.

\bibitem[Venkataraman et~al.(2016)Venkataraman, Yang, Franklin, Recht, and
  Stoica]{Venkataraman:2016:EEP:2930611.2930635}
Shivaram Venkataraman, Zongheng Yang, Michael Franklin, Benjamin Recht, and Ion
  Stoica.
\newblock Ernest: Efficient performance prediction for large-scale advanced
  analytics.
\newblock In \emph{NSDI}, 2016.

\bibitem[Yadwadkar et~al.(2016)Yadwadkar, Hariharan, Gonzalez, Smith, and
  Katz]{Yadwadkar:2017}
Neeraja~J. Yadwadkar, Bharath Hariharan, Joseph~E. Gonzalez, Burton Smith, and
  Randy Katz.
\newblock Selecting the best vm across multiple public clouds: A data-driven
  performance modeling approach.
\newblock In \emph{SOCC}, 2016.

\end{thebibliography}
\end{document}